\documentclass[journal = jacsat, manuscript = article]{achemso}
\usepackage[english]{babel}
\usepackage{graphicx}
\usepackage[utf8]{inputenc}
\usepackage{amsmath} 
\usepackage{amssymb} 
\setkeys{acs}{usetitle = true}
\setkeys{acs}{maxauthors = 0}
\usepackage[percent]{overpic} 
\usepackage[dvipsnames]{xcolor}
\usepackage{multirow}
\usepackage[normalem]{ulem}
\usepackage{bm} 

\SectionNumbersOn

\newcommand*{\kcal}{kcal$\,\,$mol$^{-1}$}

\newcommand*{\msc}{$r_\mathrm{msc}$}
\newcommand*{\euc}{$r_\mathrm{euc}$}
\newcommand*{\scc}{$r_\mathrm{scc}$}
\newcommand*{\pcc}{$r_\mathrm{pcc}$}
\newcommand*{\refwBdef}{$\omega$B97M-D3(BJ)/def2-TZVPPD}
\newcommand*{\refwB}{$\omega$B97M-D3(BJ)}

\usepackage{bm}%
\usepackage{epstopdf}
\usepackage{xfrac}
\usepackage{adjustbox}
\usepackage{tabularx}
\usepackage{booktabs}

\expandafter\ifx\csname package@font\endcsname\relax\else
 \expandafter\expandafter
 \expandafter\usepackage
 \expandafter\expandafter
 \expandafter{\csname package@font\endcsname}%
\fi
\newcommand*{\cambridge}{Yusuf Hamied Department of Chemistry, University of Cambridge, Lensfield Road, Cambridge CB2 1EW, United Kingdom}
\newcommand*{\cambridgeeng}{Engineering Laboratory, University of Cambridge, Trumpington Street, Cambridge CB2 1PZ, United Kingdom
}
\newcommand*{\ucl}{Dept. of Physics and Astronomy, University College London, 17-19 Gordon St, London WC1H 0AH, UK}
\newcommand*{\lcn}{Thomas Young Centre \& London Centre for Nanotechnology, 19 Gordon St, London WC1H 0AH, UK}

\newcommand*{\bayreuth}{University of Bayreuth, Bavarian Center for Battery Technology (BayBatt), Bayreuth, Germany}
\newcommand*{\usa}{Department of Chemical Engineering, Massachusetts Institute of Technology, Cambridge, Massachusetts 02139, United States}

\author{Philipp Pracht}\affiliation{\cambridge}\alsoaffiliation{\usa}
\email{research@philipp-pracht.de}
\author{Yuthika Pillai}\affiliation{\cambridge}
\author{Venkat Kapil}\affiliation{\ucl}\alsoaffiliation{\cambridge}\alsoaffiliation{\lcn}
\author{Gábor Csányi}\affiliation{\cambridgeeng}
\author{Nils Gönnheimer}\affiliation{\bayreuth} 
\author{Martin Vondrák}\affiliation{\bayreuth}  
\author{Johannes~T.~Margraf}\affiliation{\bayreuth}
\author{David~J.~Wales}\affiliation{\cambridge}

\phone{}

\def\inclsections{1}

\usepackage{bm}
\usepackage[hidelinks]{hyperref}
\usepackage{array}
\newcolumntype{L}[1]{>{\raggedright\arraybackslash}m{#1}} 
\newcolumntype{C}[1]{>{\centering\arraybackslash}m{#1}} 
\newcolumntype{R}[1]{>{\raggedleft\arraybackslash}m{#1}} 

\title{Efficient Composite Infrared Spectroscopy: Combining the Doubly-Harmonic Approximation with Machine Learning Potentials}

\begin{tocentry}
    \includegraphics[width=0.99\textwidth]{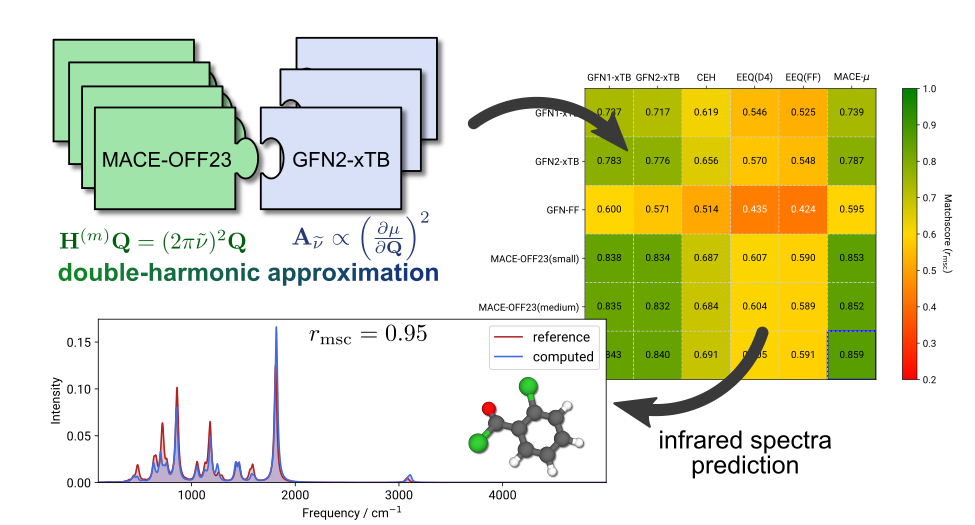}
\end{tocentry}

\begin{document}
\maketitle
\newpage

\centerline{\Large \bf Abstract}
\noindent 
\\[1cm]
{\small
{\bf \noindent Keywords:  IR spectroscopy, benchmark, machine learning potential, MACE, xTB}

\begin{abstract}
Vibrational spectroscopy is a cornerstone technique for  molecular characterization and offers an ideal target for the computational investigation  of molecular materials. 
Building on previous comprehensive assessments of efficient methods for infrared (IR) spectroscopy, this study investigates the predictive accuracy and computational efficiency of gas-phase IR spectra calculations, accessible through a combination of modern semiempirical quantum mechanical and transferable machine learning potentials.
A composite approach for IR spectra prediction based on the doubly-harmonic approximation, utilizing harmonic vibrational frequencies in combination squared derivatives of the molecular dipole moment, is employed.
This approach allows for methodical flexibility in the calculation of IR intensities from molecular dipoles and the corresponding vibrational modes. 
Various methods are systematically tested to suggest a suitable protocol with an emphasis on computational efficiency. 
Among these methods, semiempirical extended tight-binding (xTB) models, classical charge equilibrium models, and machine learning potentials trained for dipole moment prediction are assessed across a diverse dataset of organic molecules. 
We particularly focus on the recently reported machine learning potential MACE-OFF23 to address the accuracy limitations of conventional low-cost quantum mechanical and force-field methods. 
This study aims to establish a standard for the efficient computational prediction of IR spectra, facilitating the rapid and reliable identification of unknown compounds and advancing automated analytical workflows in chemistry.
\end{abstract}

\maketitle 

\ifx\inclsections\undefined
\else
\section{Introduction}
\label{sec:intro}
\fi

Infrared (IR) spectroscopy remains a key analytical tool for characterizing molecular structures and dynamics, widely applied from fundamental research to industrial quality control.\cite{bec2020,karunaratne2021,henschel2020} 
Among the various types of vibrational spectroscopy, IR spectroscopy is arguably most widely adapted and has hence been a longstanding target for computational simulations, where the two state-of-the-art approaches are  Fourier transform-based spectra prediction from both classical and quantum dynamical simulations,\cite{Luber2022} or static approaches based on the harmonic molecular Hessian and the so-called double-harmonic approximation (DHA).\cite{neugebauer2002}
While the former approach includes anharmonic effects and an averaging over conformational states via time evolution of the system, the static approach generally produces less computational overhead.
This computational efficiency can be especially important, since predicting vibrational spectra often involves time-consuming first-principles calculations that account specifically for electronic structure. 
To partially circumvent these problems, several approaches have been proposed to predict vibrational spectra from dynamical simulations via machine learning.\cite{Gastegger2017,sommers2020,Shepherd2021,kapil2024,Xu2024,Musil2022}
The static approaches have the advantage that (only) 6$N_\mathrm{atoms}$ potential evaluations 
are required to set up a seminumerical Hessian matrix, rather than several ps of molecular dynamics.\cite{Luber2022,neugebauer2002}
Unsurprisingly, the demand for quicker and more cost-effective computational approaches has led to significant developments in semiempirical quantum mechanical (SQM)\cite{elstner_chemrev,gfnWIREs} and molecular machine learning potentials (MLPs).\cite{noe2020,Butler2018,Margraf2023,deringer2021,Musil2021,Behler2021}
These two contemporary families of methodology share similar aims and designs, serving as computationally efficient alternatives to the aforementioned first-principles calculations.
Importantly, such methods are distinct from the direct deep neural network-based prediction of IR spectra,\cite{mcgill2021} and provide a more conventional integration into computational workflows. 
In this framework, advances in MLPs or SQM methods can easily be adapted, and promise greater flexibility in treating a diverse range of molecular and condensed phase systems.

Building upon our previous evaluations of semiempirical extended tight-binding and machine learning approaches for vibrational spectra,\cite{pracht2020_1,kapil2024,Musil2022} this study introduces a ``plug-in'' composite strategy (\textit{cf.} Fig.~\ref{fig:dha}) designed to enhance the speed and accuracy of IR spectral predictions and provide a snapshot of current state-of-the-art's capabilities.
By combining harmonically approximated frequencies, i.e., from diagonalization of the mass-weighted Hessian, with molecular dipole moment derivatives calculated via diverse computational techniques, the goal is to achieve a balanced approach that optimizes both computational efficiency and predictive fidelity.
\begin{figure}[ht!]
    \centering
    \includegraphics[width=1.0\linewidth]{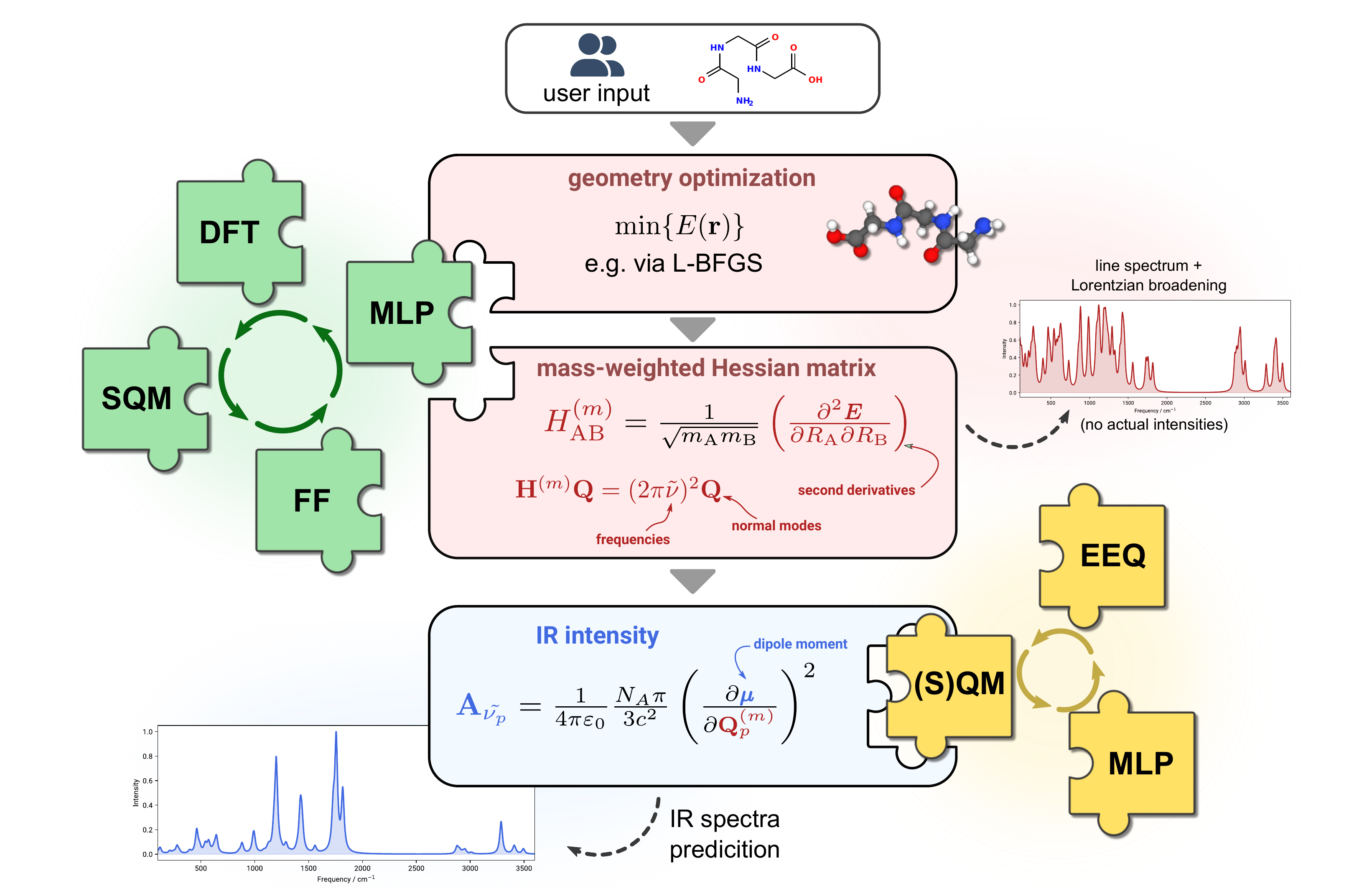}
    \caption{Composite combination of methods for IR spectra calculation within the double-harmonic approximation.}
    \label{fig:dha}
\end{figure}
Central to our methodological framework is the integration of a recently reported MLP, MACE-OFF23\cite{kovacs2023maceoff23} for the computation of molecular frequencies.
This model employs the MACE\cite{batatia2022} MLP architecture for organic molecules based on accurate 
numerical data, promising to bridge the accuracy gap typically associated with faster computational methods. 
The current implementation of this MLP does not include molecular electrostatics and so obtaining a dipole moment is not possible using only MACE-OFF23.
Therefore, a separate MACE model is introduced and trained for dipole moment prediction.
The kQEq model\cite{Staacke2022,Vondrak2023} was also considered.
We compare MLP-derived dipole moments with those obtained from classical atomic charge models and more traditional quantum mechanically derived dipole moments, providing a comprehensive evaluation of the performance for each method.
Additionally, the ability to obtain the correct molecular geometries, i.e. find the minima on the potential energy landscape via optimization procedures,\cite{wales2003book,wales2018} plays a crucial role in a composite approach, influencing both frequency calculations and dipole moment predictions.
By assessing method performance across these method capabilities, we not only seek to validate the utility of MLPs in theoretical spectroscopy, but also explore their integration into established computational workflows.
Some further key targets are to investigate whether ML methods trained on the SPICE dataset\cite{Eastman2023} are suitable surrogates for predictions at DFT-level accuracy, and whether their composite combination with other, for example SQM methods can advance our predictive capabilities in general.

With these targets in mind, we aim to advance the practical application of vibrational spectroscopy, enabling rapid and reliable identification of compounds with significantly reduced computational demands compared to conventional electronic structure calculations. 
A short summary is first provided of  IR spectra computation in the static harmonic approach and the DHA. 
The performance of the component techniques required for calculating vibrational spectra, namely geometry optimization, dipole moment, and frequency prediction, are evaluated based on a selection of benchmark sets from the literature.
Finally, IR spectra are computed putting these properties together, combining different levels of theory in a composite approach.

\ifx\inclsections\undefined
\else
\section{Theory}
\label{sec:theory}
\fi

\subsection{The double-harmonic approximation}
In the harmonic approach, molecular vibrations are derived from second derivatives of the energy $E$ with respect to the nuclear positions $R_\mathrm{A/B}$
\begin{equation}
{H}_\mathrm{AB}=\frac{\partial^{2}E}{{\partial}R_\mathrm{A}{\partial}R_\mathrm{B}} \,\,,
\end{equation}
where $\mathbf{H}$ is the Hessian matrix.
The mass-weighted force constant matrix (FCM) $\mathbf{F}^{(m)}$ is obtained from
\begin{align} \label{eq:fcm}
\mathbf{F}^{(m)} = \mathbf{M}^{-\frac{1}{2}}\mathbf{H}\mathbf{M}^{-\frac{1}{2}}
\end{align}
where $\mathbf{M}$ is the diagonal matrix of the atomic masses. 
Diagonalization of $\mathbf{F}^{(m)}$ according to
\begin{equation} \label{eq:diag_fcm}
\mathbf{F}^{(m)}\mathbf{Q}=\mathbf{\epsilon{Q}}
\end{equation}
provides a set of $3N_\mathrm{at}$ eigenvectors ${Q}^{(m)}_p$, which correspond to the normal modes in Cartesian space, and their corresponding eigenvalues $\epsilon$. 
To obtain frequencies in reciprocal centimeters $\tilde{\nu_p}$ for mode $p$, eigenvalue $\epsilon_p$ is converted with 
\begin{equation}
\tilde{\nu_p}=\frac{1}{2\pi}\sqrt{\epsilon_p}.
\end{equation}
In the so-called `double-harmonic approximation', which gets its name from the Taylor series truncation after the quadratic terms for both frequencies and dipole derivatives,\cite{porezag1996,neugebauer2002} the intensity $A_{\tilde{\nu_p}}$ for the normal mode $p$ is assumed proportional to the squared dipole derivative of the corresponding normal mode and the integral absorption coefficient is then given as
\begin{equation} \label{eq:irintensities}
A_{\tilde{\nu_p}}=\frac{1}{4\pi{\varepsilon}_0}\frac{N_A{\pi}}{3c^2}\left(\frac{\partial{\mu}}{\partial{Q^{(m)}_p}}\right)^2,
\end{equation} 
where $\frac{\partial{\mu}}{\partial{Q^{(m)}_p}}$ corresponds to the derivative of the dipole moment with respect to the normal mode coordinates, $N_A$ is Avogadro's constant, $c$ is the speed of light, and $\varepsilon_0$ is the vacuum permittivity.
In practice, the derivative of the dipole moment is evaluated with respect to the Cartesian coordinates of each atom $j$ and is afterwards transformed onto the coordinates of the vibrational normal mode $p$ according to 
\begin{equation} \label{eq:dipgrad}
\left(\frac{\partial{\mu}}{\partial{Q^{(m)}_p}}\right)^2= \sum_{\alpha} \left( \sum^{3N_\mathrm{at}}_{j} \frac{\partial{\mu_{\alpha}}}{\partial{R_j}} \frac{\partial{{R_j}}}{\partial{Q}^{(m)}_{p,j}}  \right)^2,
\end{equation}
in which $\alpha\in\left(x,y,z\right)$ are the components of the dipole moment gradient in Cartesian coordinates. 
The combination of the vibrational frequency $\tilde{\nu_p}$ of mode $p$ and the corresponding intensity in terms of the integral absorption coefficient $A_{\tilde{\nu_p}}$ determines the calculated vibrational spectrum.
The double-harmonic approximation herein leads to errors of the spectrum in both frequencies and intensities due to truncation of higher order terms of the respective Taylor series.\cite{neugebauer2002}
However, the accuracy of those two quantities can be improved independently, as discussed below.

\subsection{Computational details}

For the computation of molecular dipole moments, molecular geometries, frequencies and IR spectra, we generally refer to a benchmark set  of 7193 gas-phase molecules\cite{pracht2020_1} ranging from three to 77 atoms in size, and containing the elements H, C, N, O, F, Cl, Br, S, and P, which coincides with the elements available in MACE-OFF23.\cite{kovacs2023maceoff23}
For simplicity, we refer to this set as `IR7193' in the following.
The corresponding molecules can generally be regarded as `drug-like` with regards to the average size and chemical composition.
This set was originally composed\cite{pracht2020_1} simply by obtaining all freely available experimental gas-phase reference IR spectra from the NIST database\cite{nist} which were then compared with computed spectra for the semiempirical GFN\textit{n}-xTB approach,\cite{gfnWIREs} the force-field GFN-FF,\cite{gfnff} and DFT B3LYP-3c (an abbreviation for B3LYP-D3(BJ)$^\mathrm{ATM}$-gCP/def2-mSVP\cite{b3lyp,dftd3,dftd3bj,gcp,pbeh3c}).
B3LYP-3c herein was identified as a well-balanced DFT level of theory for IR spectra computation and can serve as a theoretical reference for testing other low-cost methods. To avoid additional conformational effects,\cite{pracht2020_1} and effects of possible experimental measurement impurities, we take the readily available theoretical gas-phase spectra calculated at B3LYP-3c level as our reference. 

Seminumerical Hessian calculations at the GFN1-xTB,\cite{gfnxtb} GFN2-xTB,\cite{gfn2xtb} and GFN-FF\cite{gfnff} levels of theory were performed with the CREST program.\cite{pracht2020,crest_new} 
The same program was used to simultaneously obtain derivatives of the molecular dipole moment with an empirical charge equilibrium (EEQ) model using parameters employed in the D4 dispersion correction, as well as dipole derivatives obtained from the so-called Charge Extended H\"uckel (CEH) model, published for the charge-adaptive q-vSZP basis set.\cite{mueller2023} 
Installation instructions for the MACE suite can be found at \href{https://github.com/ACEsuit/mace}{https://github.com/ACEsuit/mace}.
MACE-OFF23 calculations were performed using the ASE interface.\cite{ASE2017} 
The associated geometry optimizations were done with ASE's implementation of the L-BFGS algorithm and a root mean square (RMS) force convergence criterion of 1.0e$^{-4}$\,$E_\mathrm{h}/a_0$.
Seminumerical calculations of the corresponding Hessian and dipole derivatives were performed with a standalone script employing coordinate displacements of 1.0e$^{-3}$\,$a_0$. The seminumerical Hessian calculations of MACE-OFF23 were also compared with a new analytical (``autograd'') implementation.

Analysis tools for the combination of computed Hessian matrices and dipole derivatives into IR spectra can be found at \href{https://github.com/pprcht/IRtools}{https://github.com/pprcht/vibtools}.
Our comparison of IR spectra is based on a similarity `match'score \msc\ based on the Cauchy-Schwarz inequality as defined by Eq.~S3 in the supporting information
and can have values from 0 (no match) to 1 (perfect correlation).
For more information on spectra comparison with this metric and alternative measures (Euclidean norm \euc, the Spearman correlation coefficient \scc, Pearson correlation coefficient \pcc) we refer the reader to the literature,\cite{baumann1997,song2017,petrich2011,zapata2018,pracht2020_1,henschel2020,henschel2020_2} and the supplementary material.

\ifx\inclsections\undefined
\else
\section{Results and discussion}
\fi

\subsection{Performance for geometry optimization}\label{sec:geo}

The performance of a given level of theory for geometry optimization is important, because frequencies are obtained in the harmonic approximation, where the molecule must correspond to a minimum on the energy landscape. 
If the molecular structure is not a minimum, or converges to a different conformation, identified by a high root-mean-square-deviation (RMSD) of atomic coordinates,\cite{pracht2020} imaginary frequencies or incorrect fundamental modes will be observed, respectively.
Hence, having a well-defined geometry reference  is necessary.
It is generally recognized that structure optimizations at the DFT level are much less sensitive to the choice of functional than, for example, energy and property calculations.\cite{dft_best_practice}
Commonly, mGGA functionals are sufficient for geometry optimisation, where the general recommendation is to use basis sets of triple-$\zeta$ quality, or smaller basis sets in combination with corrections for the basis set superposition error (BSSE).\cite{gcp,dft_best_practice}
Reference geometries for the IR7193 set are available at the B3LYP-3c level of theory which, as a hybrid functional with BSSE-corrected basis set, is more than sufficient to compare force-field, SQM, and MLP results.
Geometry optimizations for all low-cost methods were herein started from the B3LYP-3c optimized structure in order to ensure convergence to the closest minima on the respective potential energy landscape.

The GFN1-xTB and GFN2-xTB levels of theory along with the GFN-FF  and MACE-OFF23 force-fields were used to optimize and compare structures in the database.
Tables~\ref{tab:rmsds} and \ref{tab:rmsdspercent} summarise the corresponding RMSD statistics with regard to the B3LYP-3c reference structures.
Figure~\ref{fig:rmsds} illustrates the  distribution of RMSDs as histograms compared to a  log-normal distribution.
\begin{table}
    \centering
    \begin{tabular}{l|ccc|ccc}
         &  \multicolumn{3}{c|}{\textbf{MACE-OFF23 model}} & \textbf{GFN2-xTB} & \textbf{GFN1-xTB} & \textbf{GFN-FF}\\
               & \textbf{small} & \textbf{medium} & \textbf{large} &&& \\ \hline
    \textbf{Mean}	&	0.0879	&	0.0730	&	0.0694	&	0.1254	& 0.1330 & 0.2339\\
    \textbf{Median}	&	0.0513	&	0.0478	&	0.0467	&	0.0685	& 0.0733 & 0.1219\\
    \textbf{SD}	    &	0.1162	&	0.0873	&	0.0818	&	0.1634	& 0.1852 & 0.2555\\
    \end{tabular}
    \caption{Mean, median, and standard deviation (SD) for Cartesian RMSDs calculated between the B3LYP-3c reference and MACE-OFF23(small/medium/large) and GFN{\it n}-xTB/FF optimized structures of IR7193. All values are in {\AA}ngstr\"om. Narrow distributions indicate better performance.}
    \label{tab:rmsds}
\end{table}
\begin{figure}[ht!]
    \centering
    \includegraphics[width=0.99\linewidth]{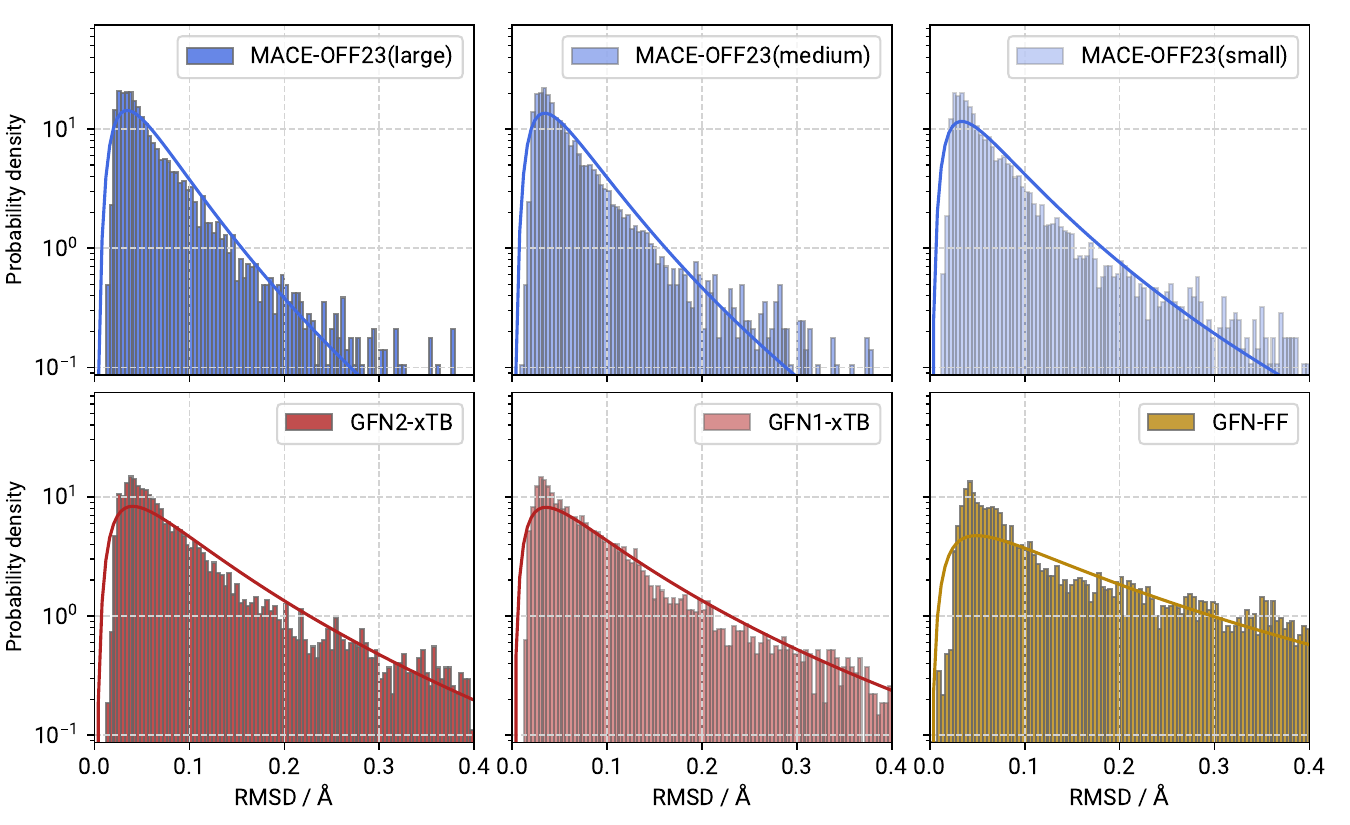}
    \caption{Histograms and fitted log-normal distributions for Cartesian RMSDs calculated between the B3LYP-3c minima and corresponding optimized molecular structures at MACE-OFF23(small/medium/large) and GFN{\it n}-xTB/FF levels of theory for the IR7193 set. All plots use a logarithmic scale to emphasize the distribution tails.}
    \label{fig:rmsds}
\end{figure}
\begin{table}[ht!]
    \centering
    \begin{tabular}{c|ccc|ccc}
         \textbf{RMSD} &  \multicolumn{3}{c|}{\textbf{MACE-OFF23 model}} & \textbf{GFN2-xTB} & \textbf{GFN1-xTB} & \textbf{GFN-FF}\\
            \textbf{[{\AA}]}   & \textbf{small} & \textbf{medium} & \textbf{large} &&& \\ \hline
$\leq$\,0.2	&	91.51\%	&	94.94\%	&	95.84\%	&	84.75\%	&	83.73\%	&	61.84\%	\\
$\geq$\,0.5	&	1.63\%	&	0.76\%	&	0.57\%	&	3.95\%	&	4.57\%	&	14.08\%	\\
$\geq$\,1.0	&	0.22\%	&	0.14\%	&	0.10\%	&	0.56\%	&	0.95\%	&	1.79\%	\\
    \end{tabular}
    \caption{Percentage of structures for IR7193 falling within the specified Cartesian RMSD threshold at a given level of theory.}
    \label{tab:rmsdspercent}
\end{table}
From the presented data we infer that MACE-OFF23 generally seems to outperform the GFN methods for geometry optimization.
In Figure~\ref{fig:rmsds} this result is shown clearly, where a narrow and sharp log-normal distribution denotes a better performance with regards to the B3LYP-3c reference.
Consequently, the average quality of geometry optimization seems to increase in the order GFN-FF, GFN1-xTB and GFN2-xTB, to the MACE-OFF23 models, which is consistent with an intuitive understanding of force-field methods being outperformed by SQM and MLP methods.
Within MACE-OFF23, the quality of the optimized geometries increases as one moves from small, to medium, to the large model. 
The mean is greater than the median because  a number of high RMSD outliers are present, which is easily confirmed by the top-heavy log-normal distributions shown in Figure~\ref{fig:rmsds}.
This difference in mean and median decreases with the quality of the MACE-OFF model used. 
Nonetheless, for all three models the mean and median of RMSDs with reference geometries is below 0.1 \AA, which is an excellent performance for this test set. 
Likewise, over 90\% of molecules in the test set have RMSDs below 0.2\,{\AA}, which illustrates the high probability of any of the three MACE-OFF models in recovering something reasonably close to the DFT geometry.
With the exception of the MACE-OFF23(small) model, less than 1\% of cases have a high RMSD over 0.5\,{\AA}, and only 16 out of 7193 cases exceed an RMSD of 1.0\,{\AA}, further attesting to the good quality of the geometries.

The GFN{\it n}-xTB methods also exhibit  excellent performance, although not as good as MACE-OFF23.
The mean RMSD of GFN1- and GFN2-xTB slightly exceeds 0.1\,{\AA} with 0.1330\,{\AA} and 0.1254\,{\AA}, respectively.
Still, with over 80\% of molecules below an RMSD of 0.2\,{\AA}, the SQM methods offer a decent quality for the optimized geometries, although there are  significantly more outliers at $\geq$0.5 and $\geq$1.0\,{\AA} RMSD compared to the MACE MLPs.
The classical force-field GFN-FF, with a mean RMSD of 0.2339\,{\AA}, is the worst performing method in this test.
Only 61.84\% of molecules are found below the 0.2\,{\AA} mark at this level of theory, which is far less than with the SQM or MLP methods.
While clearly having a computational cost advantage compared to the other tested methods, geometries are not as likely to converge to something close to the corresponding DFT minimum.

Figure~\ref{fig:rmsdexmpl} showcases some molecules that were not in good agreement with the DFT reference at either the semiempirical or MACE-OFF level of theory.
\begin{figure}[ht!]
    \centering
    \includegraphics[width=1\linewidth]{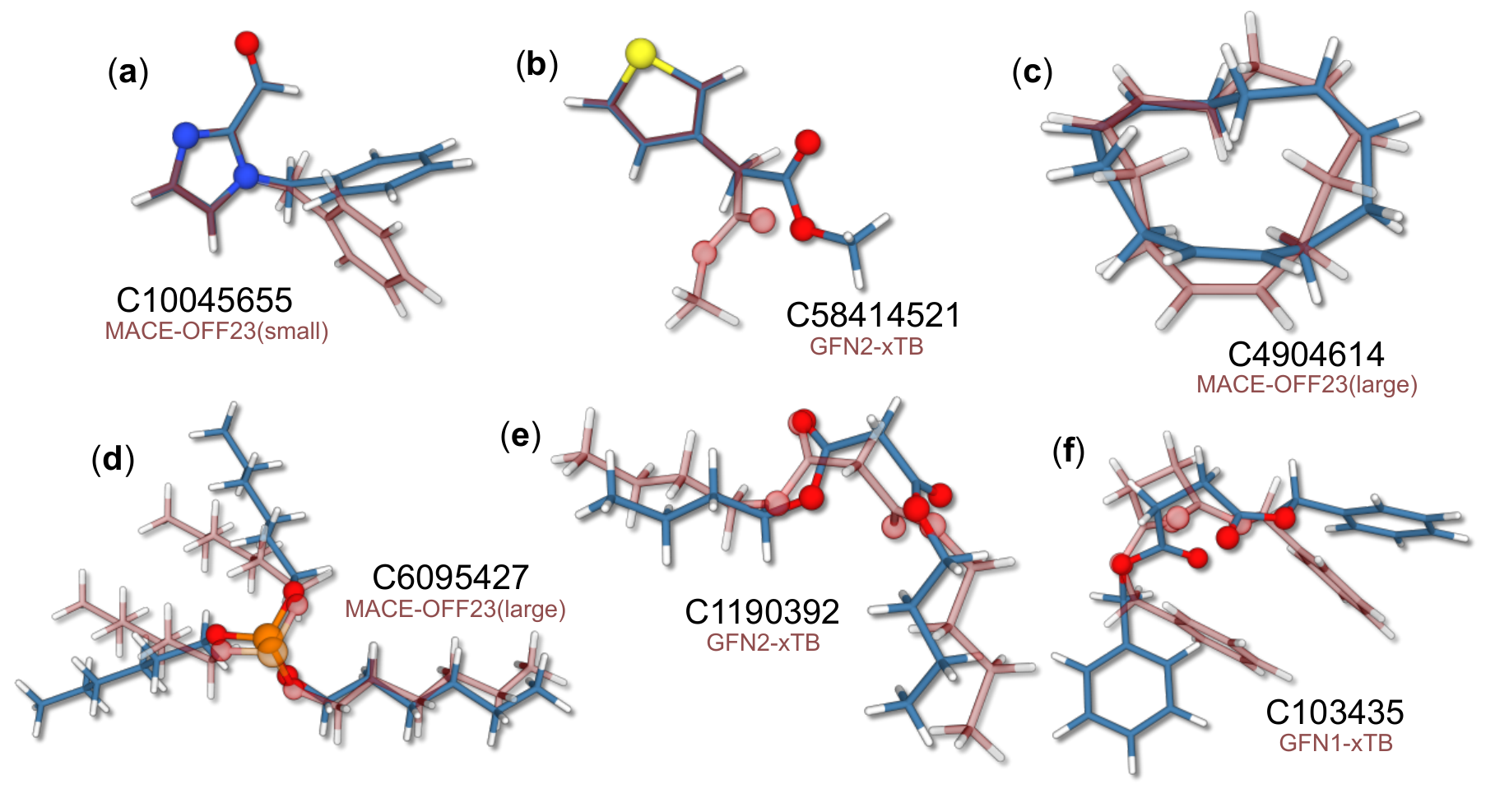}
    \caption{Selected examples for large RMSD cases of molecules from IR7193. Structures shown in blue are the B3LYP-3c reference geometry, structures in transparent red are obtained with low-cost SQM or MLP methods as indicated accordingly. NIST identifiers are given next to each structure. See text for discussion.}
    \label{fig:rmsdexmpl}
\end{figure}
Analysis of the RMSD values from different methods reveals that MACE-OFF generally performed better than the semiempirical methods, with some exceptions.
All MLP model sizes and the GFN-FF method incorrectly optimize Figure~\ref{fig:rmsdexmpl}a which shows the MACE-OFF23(small) optimized structure in comparison with the reference. 
Here, the force-field models seem unable to predict some electronic interactions and
MACE-OFF23 and GFN-FF models produce an incorrect orientation for the phenyl group, which  hints at a subtle interaction between the aldehyde hydrogen and the phenyl $\pi$-system. 

The semiempirical methods, on the other hand, account for such effects: 
the RMSD values for MACE-OFF and GFN-FF models are $\sim$1.0\,{\AA} whereas the xTB Hamiltonians produce the correct conformation and an RMSD of only $\sim$0.2\,{\AA}.  
Figure~\ref{fig:rmsdexmpl}b shows an overlay of the GFN2-xTB and reference structures.
This conformation, starting the geometry optimization from the DFT reference, is selectively produced only by the MACE-OFF23(medium) model and GFN2-xTB.
At the B3LYP-3c level, the conformational change is associated with an energy increase by 1.97\,\kcal, showing it to be a typical asymptomatic case with no clear cause for the mismatch.
A similar case is shown in Figure~\ref{fig:rmsdexmpl}c, where the MACE-OFF23(large) conformation is compared with B3LYP-3c.
All methods, except for MACE-OFF23(small), have converged to a significantly different structure from the reference structure with a high RMSD around 1.0\,{\AA}. 
However, the  conformation produced by these methods seems to be a more symmetrical minimum, possibly hinting at an incorrect reference structure. 
This is confirmed by optimization of the B3LYP-3c reference structure at $\omega$B97M-D3(BJ)/def2-TZVPP level, which converges to the same conformation as the low-cost MLP and SQM methods.

Figure~\ref{fig:rmsdexmpl}d, shows another case in which all the low-cost methods disagree with B3LYP-3c.
In fact, this molecule, trihexyl phosphite, is the worst-performing case for the MACE-OFF23(large) model, and the \textit{only} case in which an RMSD of 1.5\,{\AA} was exceeded for this level of theory.
The conformation produced by all the low-cost methods seems dominated by intramolecular dispersion interactions between two of the three hexyl chains, producing a ``hairpin''-like structure, rather than the pinwheel-like reference minimum.
Similar to Figure~\ref{fig:rmsdexmpl}c, since all methods disagree with B3LYP-3c, the error could potentially lie with the DFT reference, rather than the low-cost methods.
However, the hairpin structure is significantly less stable at the B3LYP-3c level by about 10\,\kcal. 
Furthermore, optimization of the B3LYP-3c geometry at the $\omega$B97M-D3(BJ)/def2-TZVPP maintains the same conformation, which suggests that all low-cost methods end up on a different minimum energy path leading to the hairpin minimum during the initial stages of the geometry optimization.

Two less ambiguous cases are shown in Figures~\ref{fig:rmsdexmpl}e and \ref{fig:rmsdexmpl}f, which are overlaid by GFN2-xTB and GFN1-xTB minima, respectively.
Minima for both molecules are correctly retained by the MACE-OFF23 models.
For \ref{fig:rmsdexmpl}e, the GFN methods inaccurately predict the dihedral angles between the two backbone ester groups, which seems to be a common method error in several of the molecules within IR7193.
The presence of long pentyl groups in this molecule leads to high RMSDs of 1.235 to 1.316\,{\AA}.
For \ref{fig:rmsdexmpl}f the incorrect geometry is likely a consequence of GFN1-xTB and GFN-FF overestimating the $\pi$-$\pi$ interactions between the two phenyl groups, leading to a ``sandwich''-like conformation with an RMSD of $\sim$1.25\,{\AA} for both methods.  
In summary, while some cases exist that allow a qualitative interpretation of method preference for different conformations, it does not seem possible to quantify model performance for specific intramolecular interactions based on this benchmark set, which needs to be  investigated on a by-case basis. 

\subsection{Performance for vibrational frequencies}\label{sec:vib}

Having discussed performance for optimized geometries in the previous section, we now turn our attention to the calculation of properties defining the IR spectra prediction, with a primary focus on the calculation of internal molecular vibrations in the harmonic approximation.
A simple assessment of frequency quality is possible via comparison of the harmonic zero-point vibrational energy (ZPVE), which is approximated\cite{scott1996} as half the sum over all frequencies 
\begin{equation}\label{eq:zpve}
    E_\mathrm{zpve} = \frac{1}{2}\sum_p \tilde{\nu_p} \,\,.
\end{equation}
The ZPVE together with enthalpic and entropic contributions, produces  the thermodynamic contributions to the Gibbs free energy in supramolecular calculations.\cite{grimme2021-jcpa}
Although Eq.~\ref{eq:zpve} is not precise due to neglect of anharmonicity\cite{scott1996} and a significant amount of information is condensed into the single value of the ZPVE, it offers the distinct advantage of being independent of the  permutational order of fundamental frequencies, which can vary with the level of theory used and complicate IR spectra comparison.
Furthermore, it allows for the direct derivation of frequency scaling factors, which are commonly applied in (harmonic) DFT frequency calculations to counteract the effects of inadequate descriptions of vibrational anharmonicity.\cite{scuseria1997,scott1996,pople1981,merrick2007,martin2015}

The ZPVE, calculated for all molecules in the IR7193 set at the GFN1- and GFN2-xTB, the GFN-FF, and the MACE-OFF23 levels of theory, was evaluated against reference values obtained from B3LYP-3c.
To extend the comparison, modified frequencies from a scaled version of B3LYP-3c were used to obtain linear frequency scaling factors, by taking the corresponding fraction of ZPVEs and averaging over the whole test set for each method.
The modified B3LYP-3c results, denoted as B3LYP-3c(mscal), employ a special non-linear scaling scheme following Ref.~\citenum{pracht2020_1} and  the work of Pulay et al.\cite{pulay1998,pulay1995} 
The corresponding ``mass-scaling'' factors were fitted to the experimental IR spectra and should account for a large part of the vibrational anharmonicity.\cite{irikura2005,pracht2020_1}
Using the B3LYP-3c(mscal) as a reference, the ZPVEs were evaluated twice: Once, using the raw ZPVE obtained with each low-cost potential, and once with the ZPVE of each low-cost method based on frequencies scaled by the linear factor determined from the aforementioned fraction.
Error measures (mean deviation, mean absolute error, root-mean-square error, and standard deviation) for the investigated levels of theory are shown in Table~\ref{tab:zpve}.
\begin{table}[ht!]
    \centering
    \caption{Comparison low-cost potentials for calculation of ZPVE. Three sets of data are shown: First values with plain B3LYP-3c as a reference, second B3LYP-3c mass-scaled frequencies as a reference (\textit{cf}. Ref.~\citenum{pracht2020_1}), and finally results with linear scaling factors applied and mass-scaled B3LYP-3c as a reference. Values are given in \kcal.}
    \label{tab:zpve}
    \begin{tabularx}{\textwidth}{l|XXX|XXX}
        \toprule
        \multicolumn{1}{c|}{\textbf{errors}} &  \textbf{GFN1-xTB} &  \textbf{GFN2-xTB} &  \textbf{GFN-FF} &  \multicolumn{3}{c}{\textbf{MACE-OFF23 model}} \\
        \multicolumn{1}{c|}{\textbf{[\kcal]}} &  & & &  \textbf{small} &  \textbf{medium} &  \textbf{large} \\
        \midrule
        \multicolumn{6}{l}{\textbf{raw frequencies vs. B3LYP-3c refernce}}  &  \\
        MD &  -2.46 &  -2.66 &  -5.40 &  1.15 &  1.10 &  1.09 \\
        MAE &  2.46 &  2.66 &  5.40 &  1.15 &  1.10 &  1.09 \\
        RMSE &  2.67 &  2.88 &  5.78 &  1.24 &  1.17 &  1.17 \\
        SD &  1.05 &  1.08 &  2.06 &  0.47 &  0.41 &  0.40 \\
        \midrule
        \multicolumn{6}{l}{\textbf{raw frequencies vs. B3LYP-3c(mscal) reference}}  &  \\
        MD &  1.41 &  1.21 &  -1.53 &  5.02 &  4.97 &  4.96 \\
        MAE &  1.60 &  1.51 &  1.57 &  5.02 &  4.97 &  4.96 \\
        RMSE &  2.19 &  2.14 &  1.88 &  5.54 &  5.47 &  5.46 \\
        SD &  1.68 &  1.76 &  1.90 &  2.34 &  2.28 &  2.27 \\
        \midrule
        \multicolumn{6}{l}{\textbf{scaled frequencies vs. B3LYP-3c(mscal) reference}}  &  \\
        \textbf{Scaling Factor} &  0.9913 &  0.9940 &  1.0166 &  0.9602 &  0.9606 &  0.9606 \\
        MD &  0.35 &  0.48 &  0.45 &  0.04 &  0.03 &  0.02 \\
        MAE &  1.01 &  1.15 &  1.18 &  0.23 &  0.22 &  0.23 \\
        RMSE &  1.34 &  1.57 &  1.62 &  0.31 &  0.30 &  0.30 \\
        SD &  1.29 &  1.49 &  1.56 &  0.31 &  0.30 &  0.30 \\
        \bottomrule
    \end{tabularx}
\end{table}

When using plain B3LYP-3c as a reference (topmost block in Tab.~\ref{tab:zpve}), the MACE-OFF23 model shows significantly lower errors compared to the GFN methods. 
Specifically, the MACE-OFF23 model has MAE values ranging from 1.09 to 1.15\,\kcal\ and RMSE values from 1.17 to 1.24\,\kcal, whereas the GFN methods exhibit much higher errors with MAE values between 2.46 and 5.40\,\kcal\ and RMSE values between 2.88 and 5.78\,\kcal.
However, when mass-scaled B3LYP-3c (considered here as a better ``ground truth'' for ZPVE calculations, middle block in Tab.~\ref{tab:zpve}) is used as the reference, the performance trend reverses. 
The GFN methods show improved accuracy, with MAE values between 1.51 and 1.60\,\kcal\ and RMSE values between 1.88 and 2.19\,\kcal, which are lower than the corresponding errors for the MACE-OFF23 model (MAE: 4.96 to 5.02\,\kcal, RMSE: 5.46 to 5.54\,\kcal). 
GFN2-xTB generally exhibits  slightly better accuracy compared to GFN1-xTB, while GFN-FF exhibits the largest errors among the three.
Within MACE-OF23, the small, medium, and large models exhibit very similar performance, with only minor variations in the errors, indicating consistent accuracy for harmonic ZPVE calculation by the MLP.
These results suggest that the GFN methods represent an adequate off-the-shelf solution for calculation the vibrational frequencies, while MACE-OFF23 exhibits performance closer to the plain hybrid DFT functional an will need accounting for harmonic approximation, for example via re-scaling of the corresponding frequencies.

The final block of data in Tab.~\ref{tab:zpve} provides insights into the linear scaling factors determined using the mass-scaled B3LYP-3c ZPVE as a reference.
Notably, the MACE-OFF23 models have consistent scaling factors around 0.96, which is close to the literature scaling factor of 0.97 for B3LYP.\cite{scott1996,merrick2007,martin2015}
MAEs for the MACE MLP are lowered to just 0.22 to 0.23\,\kcal, while RMSEs are lowered to 0.30 to 0.31\,\kcal.
While, to the best of our knowledge, no scaling factors have been investigated for the SPICE $\omega$B97M-D3 reference, the related $\omega$B97X-D3 functional has known scaling factors for fundamental frequencies of around 0.95,\cite{martin2015} which is close to the observed MACE-OFF23 scaling factors.
This result suggests that the MLP behaves similarly to a hybrid DFT method, supporting the previous assessment of its performance. 
Importantly, a significant harmonic frequency scaling factor shows a systematic error in particular for high-frequency vibrational modes as these dominate the harmonic ZPVE.
On the other hand, the GFN methods, being semiempirical and force-field methods, do not exhibit systematic scaling factors, reflecting the less consistent behavior in scaling vibrational frequencies that is known for these levels of theory in the literature.\cite{morokuma2004,penke2007,henschel2020} 
Nevertheless, the GFN methods MAE is lowered slightly to 1.01 to 1.118\,\kcal, and the respective RMSE is lowered to 1.34 to 1.62\,\kcal.
Overall, the consistency in the MACE models underscores their reliability and further validates their use for molecular frequency calculations within the harmonic approximation, aligning closely with hybrid DFT methodologies.
Nonetheless, the GFN methods may have an advantage in computational efficiency for drug sized molecules, as addressed in the following.

\subsubsection{Computational efficiency for harmonic frequencies}

Efficiency is a major concern for the high-throughput \textit{in silico} analysis of chemical systems, both in terms of computational cost and in terms of time to solution. In the context of IR spectra, the main computational bottleneck is the computation of the Hessian matrix. For ML potentials, forces are usually obtained via automatic differentiation (autograd). This involves constructing a computational graph of all operations required for the energy calculation and subsequently computing the gradients via backpropagation. Using forces of displaced geometries, (semi)numerical Hessians can be computed, as described above. A second pass through the graph also allows for the computation of second derivatives with respect to the atomic positions, yielding the elements of the Hessian matrix directly. This offers a more efficient and precise way to calculate derivatives. Figure~\ref{fig:timings} shows MACE timings for Hessian calculations using the numerical and autograd implementations as a function of molecule size.

\begin{figure}
    \centering
    \includegraphics[width=0.99\linewidth]{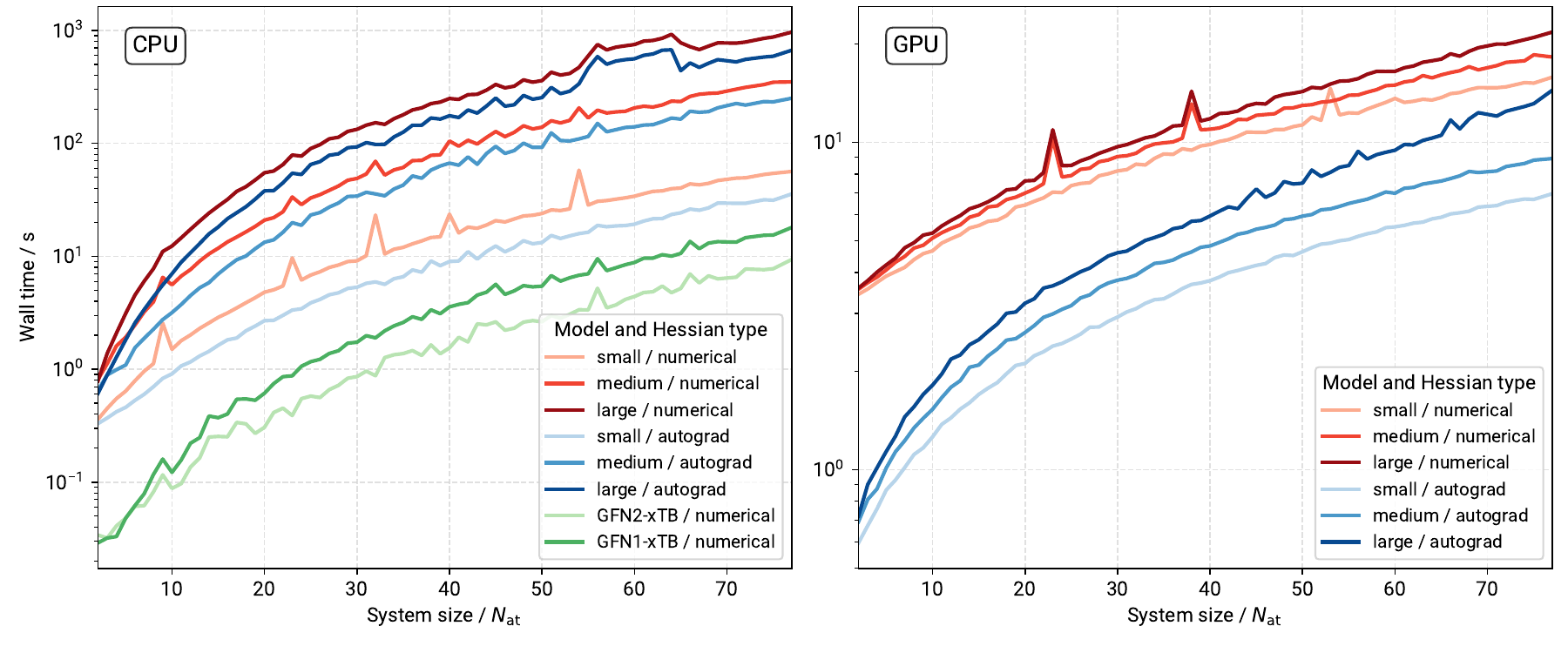}
    \caption{Wall-time comparison for MACE-OFF23 models and Hessian matrix calculations on CPU (left) and GPU (right) devices. Hessian matrices were calculated either by a numerical two-sided difference procedure from the molecular gradients (red), or via auto differentiation (blue). $N_\mathrm{at}$ is the system size (number of atoms). The wall times are given in seconds on a logarithmic scale. All CPU calculations used 4 threads for shared memory (OpenMP) parallelization and were obtained on a 11th Gen Intel Core i7-11800H (2.30GHz) processor. The CPU implementation was used here to ensure direct comparability to the xTB calculations.}
    \label{fig:timings}
\end{figure}

On CPUs (using four shared memory threads), the MACE MLPs are significantly slower than the semiempirical xTB methods for the relatively small molecules investigated herein.
For example, the largest molecule in IR7193, with 77 atoms, takes roughly 35 seconds for a Hessian calculation with the efficient autograd implementation and the smalles MACE-OFF23.
The same molecule, using the same computational resources, is processed with a numerical Hessian at the GFN2-xTB level in just 8 seconds. Similar timings can be achieved, when the MACE models are evaluated using GPU resources, however. On balance, this means that the xTB methods have an edge in terms of computational efficiency in high-throughput applications for drug-like molecules, when only CPU resources are available and xTB is sufficiently accurate. 

However, this conclusion is valid only for small systems because the rate-determining step in an xTB calculation formally scales as $\mathcal{O}(N^3)$ (with $N$ being the system size with regards to the number of basis functions) which is the cost for diagonalization of the Hamiltonian.
A prefactor for this scaling comes from the usually high number of self-consistent charge (SCC) iteration cycles required to converge to a solution for large systems.
At system sizes of around seven to eight hundred atoms each xTB energy evaluation is roughly the same cost as a MACE-OFF23(medium) evaluation and going beyond that MACE quickly becomes the much cheaper option, even on CPUs. The same ``turn-over point'' for the MACE-OFF23(small) models is located even earlier at around two to three hundred atoms (\textit{cf.} Supporting Information).
Furthermore, xTB may not be able to converge to a suitable SCC solution for large systems at all, experiencing similar self-consistent field issues to a DFT method.
On the other hand, no MACE-OFF23(large) calculation was faster than any equivalent xTB evaluation on CPUs. Furthermore, the large MACE model has more extensive memory requirements than either the xTB methods or the small and medium models which one needs to consider when performing calculations.
In our opinion, the use of the MACE-OFF small and medium MLPs could therefore be beneficial for large amorphous (e.g. peptide and protein) or condensed phase systems, as these models provide the best cost to accuracy ratio.
Future work with a thorough investigation of such systems is planned by our groups.

When GPU resources are available, the MACE models are overall the best choice, both in terms of time-to-solution and in terms of accuracy (see below). Notably, the benefit of the autograd implementation is even more significant on GPUs, with autograd Hessians being cheaper for the large model than numerical ones for the small one, across the full investigated size range.

\subsection{Performance for molecular dipole moments}\label{sec:dipoles}

IR intensities are related to the molecular dipole moment $\mu$, as described by Eq.~\ref{eq:irintensities}.
Therefore, if attempting to switch out the method used for the calculation of $\partial \mu/\partial Q_p$ in a composite approach, it is sensible to benchmark the performance for $\mu$.
Employing again the IR7193 set of molecules in addition to three conformational benchmark sets (MALT222,\cite{glucose-maltose} MPCONF196,\cite{rezac2018} and 37conf8\cite{shapara2019}), this comparison was done for the B3LYP-3c level of DFT, the semiempirical GFN1- and GFN2-xTB Hamiltonians, the CEH method,\cite{mueller2023} an empirical EEQ model using parameter sets from the D4 dispersion correction and the GFN-FF force-field, as well as a novel MLP for dipole prediction, {MACE-$\mu$}, and a new fit of the kQEq model\cite{Staacke2022,Vondrak2023} for molecules containing the elements HCNOF.
Both MACE-$\mu$ and kQEq were trained on the SPICE dataset\cite{Eastman2023} for this study.
The \refwBdef\ level of theory was used as a reference for these dipole calculations, and is also used as the reference in the SPICE dataset.
The \refwB\ level itself was benchmarked for dipole prediction on a much smaller set of molecules, originally composed by Head-Gordon et al.\cite{mhg2018,Zapata2020} and refers to  CCSD(T)/CBS data, which can be found in the supporting information.

Since IR intensities are calculated from \textit{derivatives} of the dipole moment (Eq.~\ref{eq:dipgrad}) rather than the dipole moment directly, two criteria are important for the prediction of $\mu$:
first, the total dipole moment, and second the orientation of the dipole vector.
The total dipole moment is straightforward to benchmark, where we employ  the regularized dipole metric $\Delta\mu = {|\mu-\mu_{\mathrm{ref}}|}/{\max{\{\mu_{\mathrm{ref}}},1~\mathrm{Debye}\}}$ proposed by Hait and Head-Gordon,\cite{mhg2018} in order to handle small and large dipole moments on a more even footing. 
The corresponding results are shown as violin plots for the summarized IR7193, MALT222, MPCONF196 and 37conf8 sets in Fig.~\ref{fig:dipolviolin} and Table~\ref{tab:dipole}.
Additional information is provided in the supporting information.
\begin{figure}[ht!]
    \centering
    \includegraphics[width=0.99\linewidth]{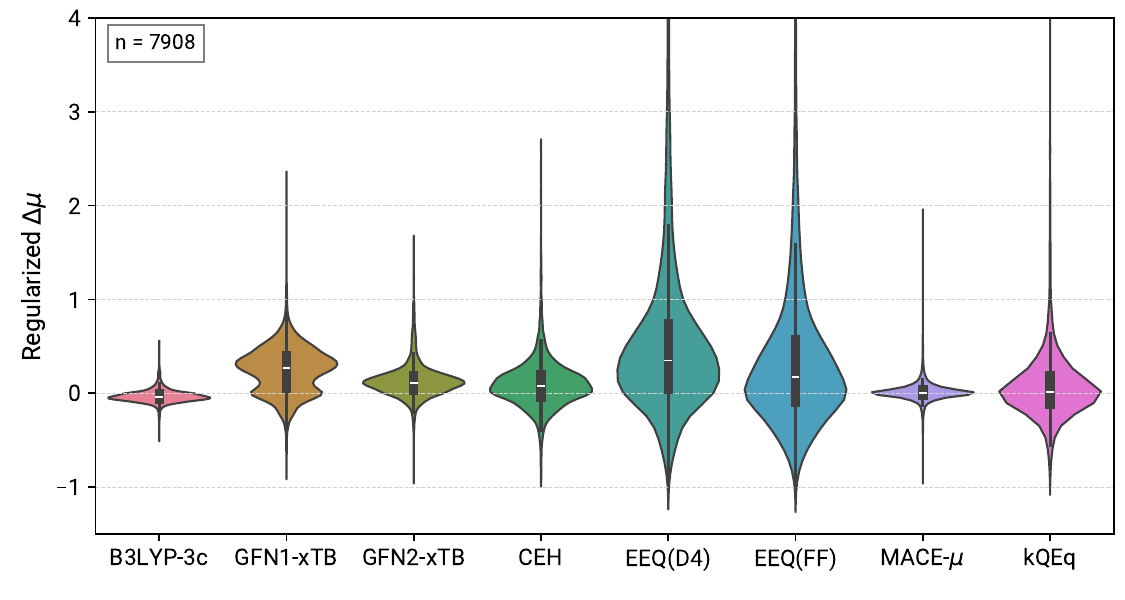}
    \caption{Violin plots for regularized errors of dipole moments at a given level of theory with regard to the reference $\omega$B97M-D3(BJ) dipole moments. Note, kQEq was evaluated for the subset of molecules containing HCNOF elements only. }
    \label{fig:dipolviolin}
\end{figure}
\begin{table}[ht!]
    \centering
    \begin{tabularx}{\textwidth}{lXXXXXXXX}\hline
 & B3LYP-3c & GFN1-xTB & GFN2-xTB & CEH & EEQ(D4) & EEQ(FF) & MACE-$\mu$ & kQEq \\ \hline
MD & -0.0325 & 0.2443 & 0.1320 & 0.0941 & 0.5424 & 0.3620 & 0.0128 & 0.0754 \\
MAD & 0.0480 & 0.1944 & 0.1213 & 0.1723 & 0.5680 & 0.5451 & 0.0565 & 0.2331 \\
RMSE & 0.0697 & 0.2449 & 0.1877 & 0.2488 & 0.8813 & 0.8351 & 0.0931 & 0.4177 \\
\hline
\end{tabularx}
    \caption{Errors for regularized dipole moments at a given level of theory. Reference dipole moments were obtained at the $\omega$B97M-D3(BJ) level. Note, kQEq was evaluated for the subset of molecules containing HCNOF elements only. All values are dimensionless. }
    \label{tab:dipole}
\end{table}
We chose to evaluate the dipole orientation as the dot product of the normalized dipole vectors $\hat{\mu}\cdot\hat{\mu}^\mathrm{ref}$ of the calculated and reference dipoles, which provides a simple but meaningful way to interpret alignment. 
Values for this measure can technically be in the range of -1 to 1, where the latter indicates perfect alignment, and values with a negative sign show opposite orientation of the dipole vector.
Normal distributions of this measure were fitted for IR7193 and are shown in Fig.~\ref{fig:dipolorient}.
\begin{figure}[ht!] 
    \centering
    \includegraphics[width=0.49\linewidth]{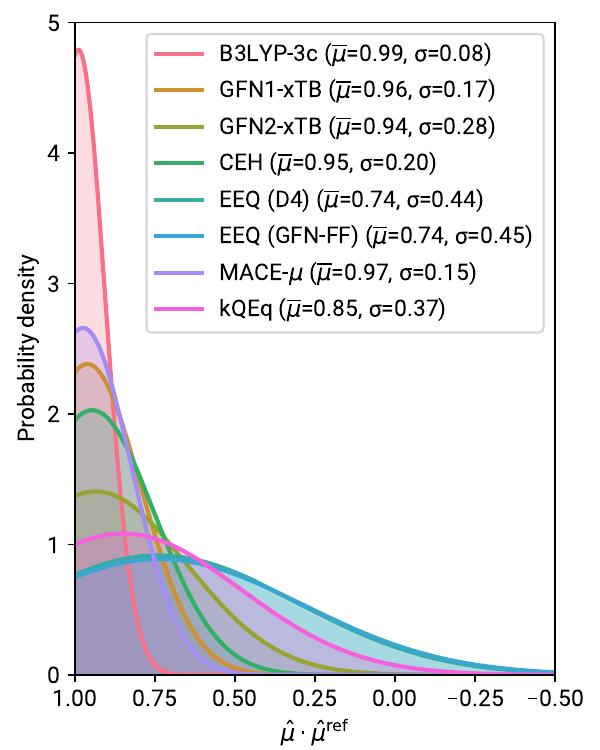}
    \caption{Normal distribution for the dot products of normalized dipole moments at a given level of theory $\hat{\mu}$ and the reference $\omega$B97M-D3(BJ) dipole moments $\hat{\mu}^\mathrm{ref}$. The respective mean $\overline{\mu}$ and standard deviation $\sigma$ are given in the legend.}
    \label{fig:dipolorient}
\end{figure} 

Several observations can be made for the dipole moment calculations. 
B3LYP-3c, which serves as our reference for IR spectra, molecular geometries and ZPVE, shows performance very close to the \refwB\ reference with a regularized RMSE of just 6.97\%.
Furthermore, as can be seen from Fig.~\ref{fig:dipolorient}, B3LYP-3c and the high-level DFT reference also exhibit  the best dipole vector alignment with a mean close to unity and a standard deviation of just 0.08.
Dipole moments at this level of theory on average seem underestimated, leading to the MD of -3.25\%.
A probable origin of this error is the small def2-mSVP basis set employed here.\cite{Zapata2020}

GFN1-xTB and the CEH model dipole moments are based on the distribution of Mulliken atomic charges.\cite{gfnxtb,mueller2023}
Unsurprisingly, both methods exhibit similar performance for total dipole moments, with a regularized RMSE of 0.2449 and 0.2488, respectively.
Characteristically, GFN1-xTB has two maxima in the violin probability density of Figure~\ref{fig:dipolviolin}, resulting in an up-shifted mean compared to CEH and a significant number of molecules have strongly overestimated dipole moments.
Both methods contain several outliers, including values overestimated by more than a factor of two according to the regularized dipole moments.
The alignment of dipole vectors of both methods with the reference is likewise similar, although significantly worse than B3LYP-3c.
Overall, GFN1-xTB slightly outperforms CEH. 
This result can likely be attributed to the SCC procedure in GFN1-xTB allowing a better adaptation of charge density, while CEH is a ``single-shot'' diagonalization (H\"uckel) method, lacking self-consistency.

GFN2-xTB improves upon both Mulliken charge based models by atom centered dipole and multipole moments.\cite{gfn2xtb}
While this description leads to a  narrower probability distribution and better mean in Figure~\ref{fig:dipolviolin} with a RMSE of 0.1877, dipole alignment is on average worse than with either GFN1-xTB or CEH.
Taking this result into account, all SQM methods offer similar robust and efficient dipole moment calculations.

MACE-$\mu$, a MACE model trained on the SPICE\cite{Eastman2023} dataset to predict the molecular dipole vector for this work, shows some interesting features. 
The model presented here formally refers to a ``medium'' MACE model size, some additional data for a ``small'' model can be found in the Supporting Information.
The MD of 0.0128, the MAD of 0.0565 and RMSD of 0.0931, as well as the dipole vector alignment (\textit{cf}. Fig.~\ref{fig:dipolorient}), show a generally better performance of MACE-$\mu$ than those of the SQM-based methods.
Instead, the MACE-$\mu$ performance much more closely resembles that of a DFT method, although some greater outliers can be seen in the violin plots of Figure~\ref{fig:dipolviolin}. 
Notably, our dipole moment reference method is the same as the level of theory employed in the SPICE dataset and, in fact, a number of predictions match the $\omega$B97M-D3 reference precisely. 
At the present time it is unclear what causes the remaining dipole moment outliers of MACE-$\mu$, but still we see it as a slightly better (although for small systems more expensive) alternative to the GFN{\it n}-xTB dipole predictions.

The worst molecular dipole moments come from the classical EEQ models. The RMSE, 0.8812 and 0.8351 for the D4 and GFN-FF parameter set, respectively, is more than an order of magnitude worse than for B3LYP-3c, and roughly a factor of four worse than with the SQM methods.
This is in line with previous observations for dipole moments calculated from EEQ models in the literature.\cite{Staacke2022,pracht2020_1}
The performance can apparently be influenced by the parametrization, and the GFN-FF EEQ model, on average, shows marginally better results than the D4 EEQ parameter set.
Unsurprisingly, the largest outlier, with a dipole moment overestimated by roughly a factor 10, is obtained with EEQ(D4).
For the dipole vector orientation both EEQ parameter sets perform equally badly, giving  broad distributions in Figure~\ref{fig:dipolorient} that extend even into the negative regime of $\hat{\mu}\cdot\hat{\mu}^\mathrm{ref}$.
Hence, a non-negligible number of molecules have dipole moments pointing into the wrong direction at this level of theory, two of which are shown in Figures~\ref{fig:dipole_examples}a and \ref{fig:dipole_examples}b.
In these cases, EEQ models clearly provide an insufficient description of the charge distribution, which can only be recovered with electronic structure calculations or by certain MLPs.\cite{Staacke2022}
EEQ models are prone to unphysical charge-transfer problems,\cite{Ko2021,Staacke2022,gfnWIREs} which contributes to these issues.

Interestingly, the kQEq model (which is an ML augmented EEQ model) performs significantly better than its classical counterparts. In terms of the dipole moments it approaches the accuracy of GFN1-xTB, although several significant outliers lead to a somewhat larger RMSE. The dipole orientations are also improved relative to the classical models, although a few outliers with dipoles pointing in the wrong direction remain. Overall, environment dependent electronegativities, as used in kQEq,  improve the performance of EEQ-based models, but the underlying limitations (\emph{i.e.} using a point-charge representation of the density and unphysical charge transfer effects of EEQ) are not fully cured.   

\begin{figure}[ht!]
    \centering
    \includegraphics[width=1\linewidth]{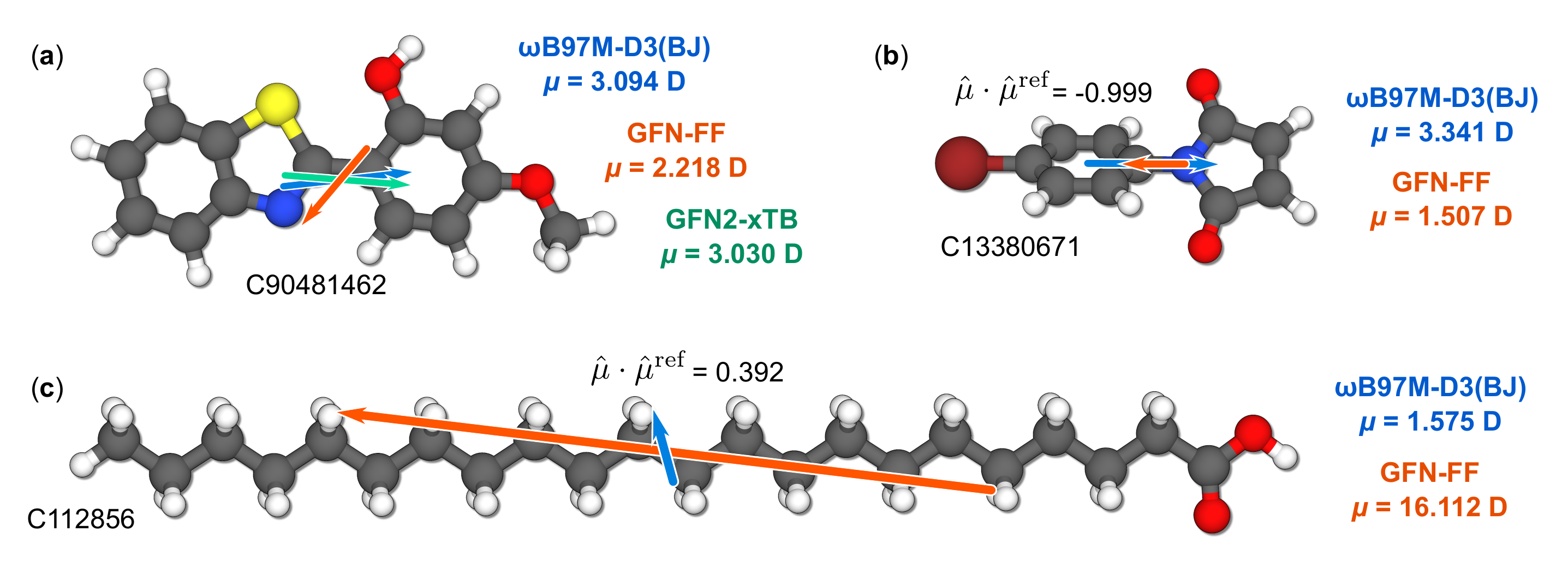}
    \caption{Example molecules with depicted dipole moments. \textbf{a}) Benzothiazolyl methoxyphenol (NIST Id C90481462), \textbf{b}) Bromophenyl maleimide (NIST Id C13380671), \textbf{c}) Docosanoic acid (NIST Id C112856). Dipole moments are depicted to-scale as colored arrows, where blue refers to \refwB, red-orange refers to EEQ(GFN-FF) and green refers to GFN2-xTB. Debye values are also included. Figures \textbf{b} and \textbf{c} furthermore illustrate the alignment measure $\hat{\mu}\cdot\hat{\mu}^\mathrm{ref}$ for the two depicted levels of theory.}
    \label{fig:dipole_examples}
\end{figure}

Figure~\ref{fig:dipole_examples}c shows the aforementioned high EEQ dipole moment outlier. A similar linear molecule has been discussed, for example, in Ref.~\citenum{Staacke2022}.
Interestingly, this outlier cannot be attributed to the partial charge distribution of the carbon atoms, as illustrated by Figure~S3 in the supporting information.
In fact, all atomic charges at the EEQ(GFN-FF) level are very similar to those at the GFN2-xTB level, despite the latter calculation exhibiting  a qualitatively well-defined dipole moment that differs from the high-level DFT reference by just 0.204\,D (or 13\%, referring to the regularized dipole measure). 
On the other hand, partial charges at the EEQ(D4) level are different from both GFN2-xTB and EEQ(GFN-FF), as well as the DFT references.
A clearly wrong assignment of charges that would be reflected in   the dipole moment prediction cannot be discerned in either case, although partial charges calculated at the high quality reference \refwB\ differ distinctively from all other methods.
The severely overestimated dipole moments via the EEQ models can therefore not be unequivocally attributed to charge transfer issues from polarization or electron delocalization.
A delicate balance between different contributions to the molecular dipole moment is needed for useful performance.

\subsection{Performance for simulation of gas-phase IR spectra}

Finally, combining predictions of harmonic frequencies and Cartesian dipole derivatives via Eq.~\ref{eq:dipgrad}, allows the prediction of IR spectra.
In principle, any combination of methods for frequencies and dipole moments can be used in this approach. 
However, both quantities are required for the same stationary point, meaning that dipole moment derivatives in particular must be obtained for molecular structures optimized at the same level of theory that is employed for the frequency calculation.

For the IR7193 set, we investigated all combinations of the discussed levels of theory. Two sets of reference data were employed: unscaled B3LYP-3c and the respective mass-scaled results, which closely correspond to experimental data.\cite{pracht2020_1} For brevity, the following discussion focuses on the \msc\ matchscore metric, with other metrics provided in the Supporting Information. Results are summarized in Figures \ref{fig:heatmap} to \ref{fig:heatmap3}, illustrating the average \msc\ over IR7193 for all method combinations, referencing different frequency scaling for either reference or actual data.
\begin{figure}[ht!]
    \centering
    \includegraphics[width=1\linewidth]{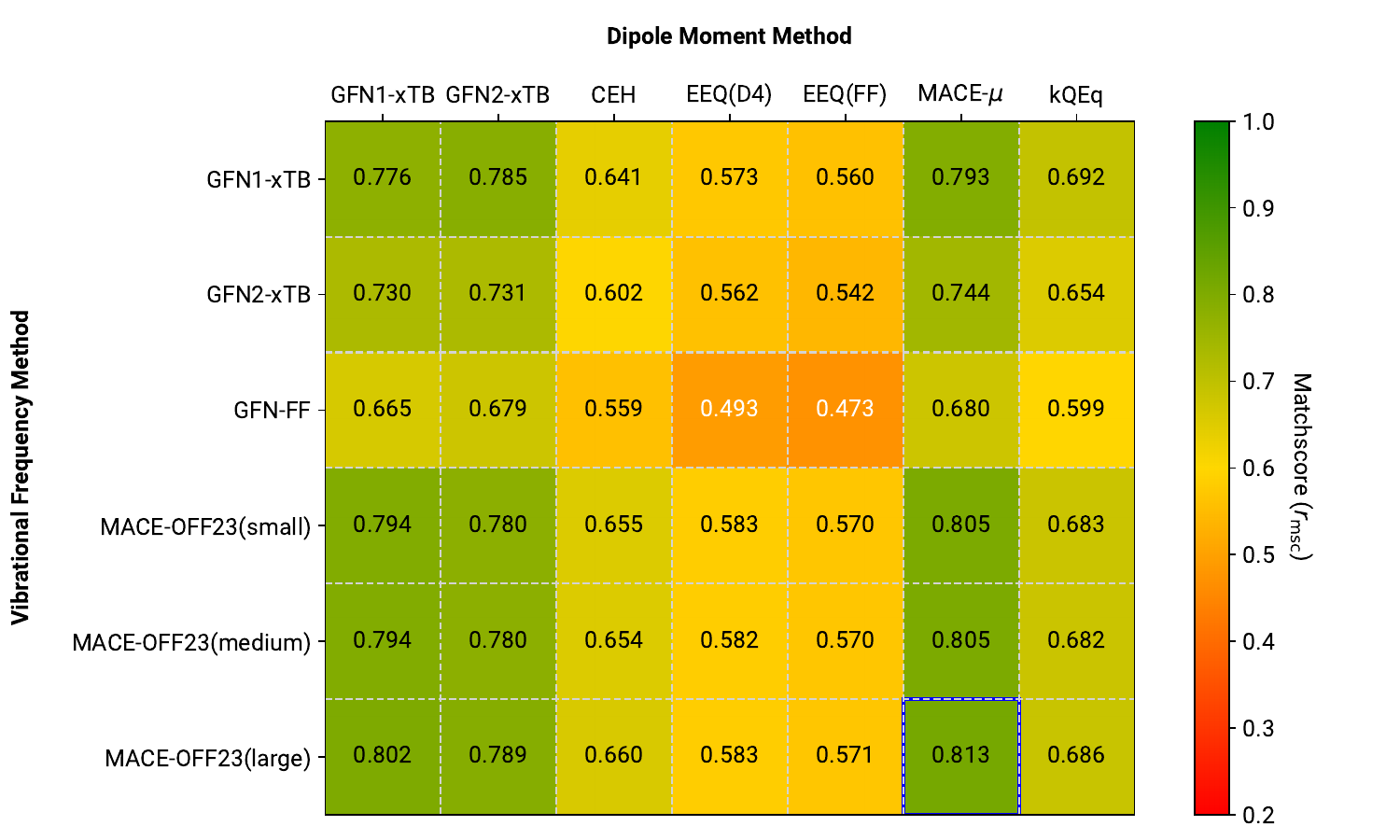}
    \caption{Average \msc\ comparison for different frequency/dipole method combinations. Reference data refers to ``plain'' B3LYP-3c IR spectra of the IR7193 set. Neither the tested methods, nor the reference, employ any kind of scaling to the harmonic frequencies. The overall highest average matchscore is marked by a blue outline.}
    \label{fig:heatmap}
\end{figure}
Figure \ref{fig:heatmap} presents the average \msc\ with respect to unscaled B3LYP-3c reference spectra. 
As expected, there is a correlation between previously observed performance for frequencies, dipole moment prediction, and the similarity scores for IR spectra. 
All MACE-OFF23 variants consistently outperform the GFN{\it n}-xTB methods, regardless of the origin of the  dipole data employed. 
The best average \msc, with a value of 0.813, is provided by MACE-OFF23(large) in combination with MACE-$\mu$ based IR intensities. 
Combinations of MACE-OFF23 with dipole derivatives obtained from either of the xTB methods also show high \msc\ values, around 0.79.
Unfortunately, IR intensities obtained via the classical EEQ model show much poorer performance, reflecting the previously observed lack of accuracy in dipole moment prediction. 
The semi-classical CEH model performs between the xTB and EEQ-derived IR predictions, contrary to the similar behavior of this method for dipole moments compared to GFN1-xTB (see section~\ref{sec:dipoles}). 
A comparable performance, moderately improving on CEH, is observed for IR spectra with intensities predicted by the kQEq method.
As observed for the dipole moments above, kQEq significantly improves upon the classical EEQ models, but the underlying methodical limitations prevent reaching the accuracy of a self-consistent semiempirical treatment. 
However, note again that kEQq was trained only for the HCNOF element subset and matchscore averages are not fully comparable to the other methods.
In general, similar trends are observed for the combination of GFN{\it n}-xTB frequencies with various dipole moment methods: MACE-$\mu$ and xTB-derived IR intensities are accurate, while classical and CEH-based IR intensities are not.
Employing a force-field, even one trained for frequency prediction such as GFN-FF, notably worsens the obtained IR spectra. 
Multiple factors probably contribute to this performance. Most importantly, GFN-FF has a significant number of large RMSD minima after optimisation, as discussed in section~\ref{sec:geo}, which naturally affects the predicted frequencies and dipoles and hence must contribute to the poor results.
The best predictions are again obtained in combination with GFN{\it n}-xTB or MACE-$\mu$ calculated IR intensities, although the assocuated computational costs exceed those for calculating the GFN-FF Hessian.

Employing mass-scaled reference data to better reproduce experimental gas-phase spectra\cite{pracht2020_1} has a significant impact on the observed performance, as shown in Figure~\ref{fig:heatmap2}.
\begin{figure}[ht!]
    \centering
    \includegraphics[width=1\linewidth]{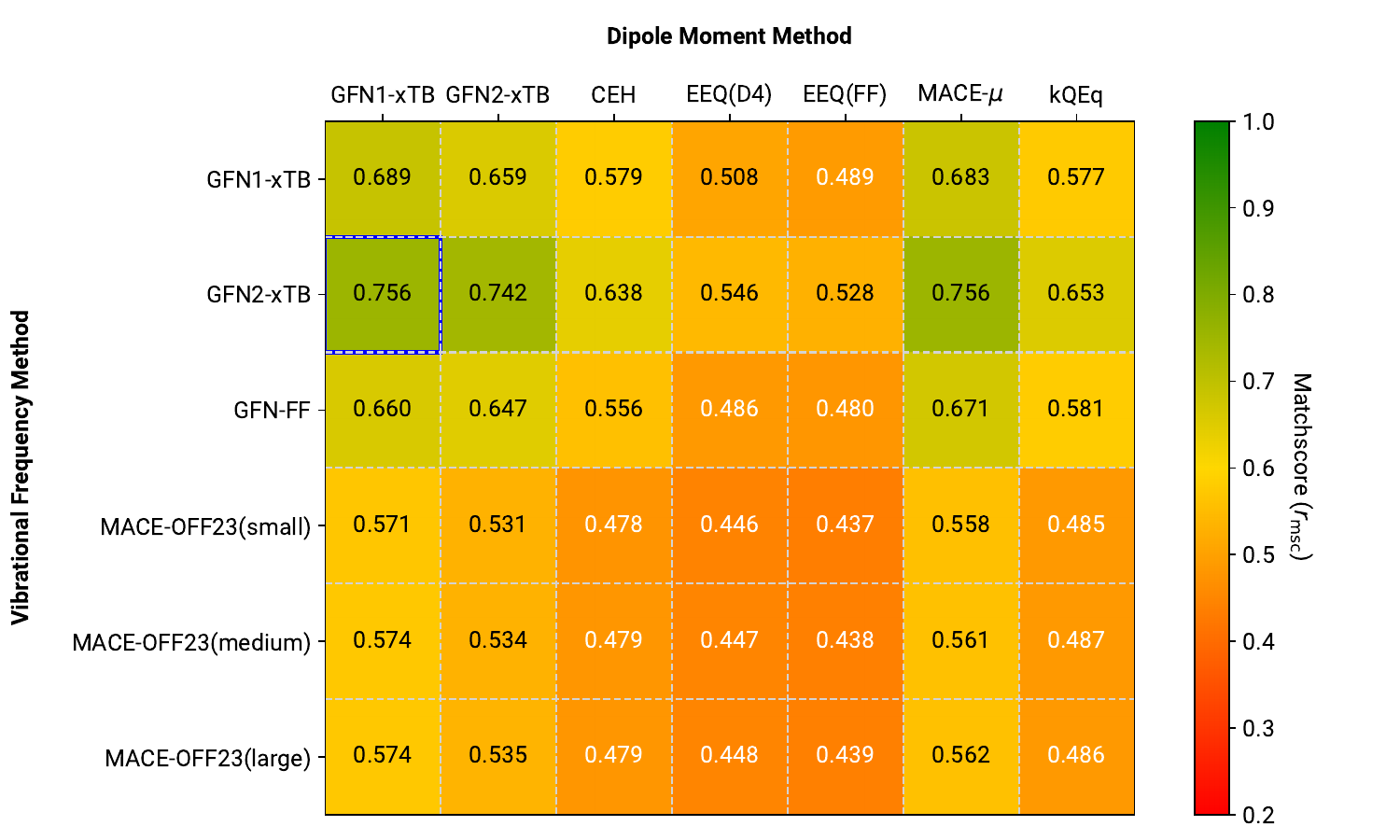}
    \caption{ Average \msc\ comparison for different frequency/dipole method combinations. Reference data refers to mass-scaled B3LYP-3c IR spectra of the IR7193 set.  Neither of the tested methods employs any kind of scaling to the harmonic frequencies. The overall highest average matchscore is marked by a blue outline.}
    \label{fig:heatmap2}
\end{figure}
All MACE-OFF23-based calculations show substantially poorer \msc\ values, regardless of the employed dipole moment methods, which can be clearly assigned to the shift in frequencies.
On the other hand, all GFN methods are much less affected by the changing reference.
GFN2-xTB  performs better than its predecessor GFN1-xTB, and even has slightly higher \msc\ than with the plain B3LYP-3c reference.
Apparently, the formulation of the GFN method family is more consistent in reproducing the molecular frequencies (as observed already for the ZPVE, see section~\ref{sec:vib}), and hence preform similarly in both cases.
This result  shows that the GFN methods are a reasonable choice for inexpensively calculating frequencies, since, on average, robust performance is expected.
Unfortunately, this observation also implies a general inability of the GFN methods to achieve much better IR spectra predictions, as well as potential shortcomings of the comparison metric, which is discussed below.

In Figure~\ref{fig:heatmap3} both the reference data and the low-cost simulations target IR spectra closely corresponding to the experiment.\cite{pracht2020_1}
\begin{figure}[ht!]
    \centering
    \includegraphics[width=1\linewidth]{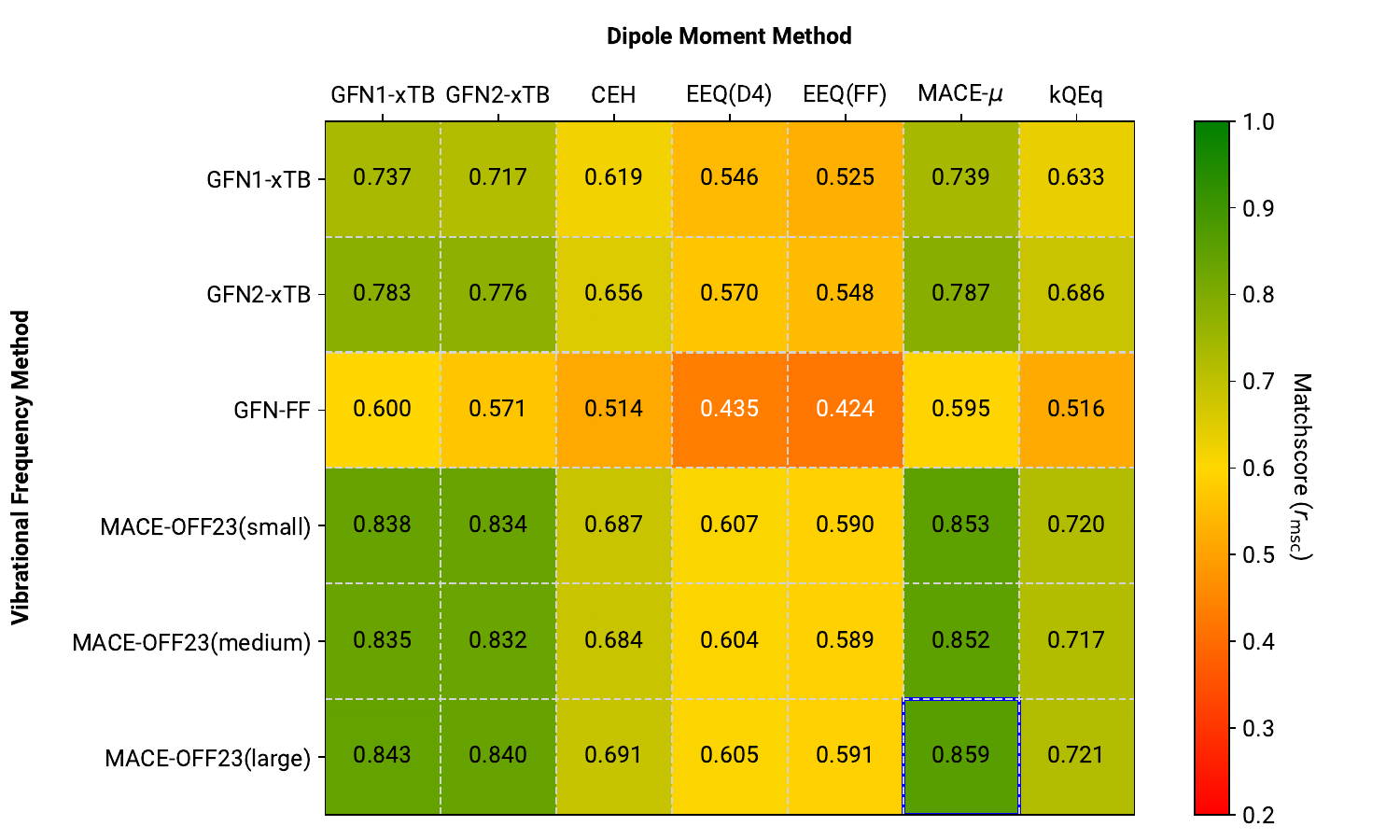}
    \caption{Average \msc\ comparison for different frequency/dipole method combinations. Reference data refers to mass-scaled B3LYP-3c IR spectra of the IR7193 set. Harmonic frequencies from the low-cost methods were scaled by the factors from Tab.~\ref{tab:zpve}. The overall highest average matchscore is marked by a blue outline.}
    \label{fig:heatmap3}
\end{figure}
Compared to Figure~\ref{fig:heatmap2}, the behaviour of MACE-OFF23-based predictions with regard to the mass-scaled B3LYP-3c reference can be rectified by employing the linear frequency scaling factors obtained in section~\ref{sec:vib}.
In combination with either the xTB-based dipole derivatives or MACE-$\mu$, all MACE-OFF23 MLPs consistently achieve average \msc\ over 0.8, with the overall best benchmark of 0.859 for MACE-OFF23(large)+MACE-$\mu$.
A similar trend is seen for GFN1-xTB and GFN2-xTB, with the latter theory performing better.
The best average \msc\ among these SQM methods is 0.787, which is achieved with GFN2-xTB in combination with MACE-$\mu$.
Combining the MACE-OFF23 frequencies with kQEq-based IR intensities generally passes a \msc\ of 0.71, which makes them roughly comparable to GFN1-xTB IR spectra employing GFN2-xTB dipole derivatives, moderately improving upon the CEH results.
As before, GFN-FF-based predictions of IR spectra, as well as all IR spectra calculated with underlying EEQ or CEH dipole moment derivatives, are lacking accuracy.
In fact, GFN-FF results scaled by the linear factor determined via the ZPVE are  on average worse than both these alternatives, as seen in Figure~\ref{fig:heatmap} and Figure~\ref{fig:heatmap2}, which is  further testimony to an asystematic behaviour of the method.
For the underlying dipole moment derivatives, classical charge-equilibrium models exhibit relatively poor average \msc\ values throughout, independent of both reference values and methods used for frequency calculation.
Hence, their use in IR spectra computation should generally be avoided.
The semiclassical CEH model, on the other hand, while certainly worse than either a SQM or specially trained MLP, shows promising performance.
Results for this method could certainly be improved with a specialized parametrization, and could motivate further development of similar single-diagonalization SQM Hamiltonians.

Unfortunately, these results can also be interpreted as a lack of precision in matchscore-based metrics like \msc. 
While clearly defined or linearly shifted spectra are sufficiently well compared by matchscore metrics, for asystematically behaving methods the comparison is too insensitive to accurately distinguish spectra, and will produce some average \msc\ value (as seen for GFN1-xTB and GFN2-xTB, as well as their combination with MACE-$\mu$ in Figure~\ref{fig:heatmap2}). 
A better metric should include clear assignment of vibrational modes between reference and calculated spectra, for which, to the best of our knowledge, no automated procedure exists, although attempts exist to project the IR intensity along localized modes and use an appropriate assignment accordingly.\cite{reiher2009vib2}
Furthermore, such a metric would only work if it is known \textit{a priori} that the same conformation of the same molecule is to be compared and information about the vibrational modes is accessible for the reference spectra.
If the goal is, for example, the identification of unknown compounds from experimental measurements, only matchscore metrics seem suitable. 
Hence, due to lack of better alternatives, we continue to refer to \msc, but advise caution to the reader. 
On the other hand, obtaining generally mediocre values of \msc\ as seen for GFN{\it n}-xTB, may signify suitability in applications where an exact order of the frequencies and modes is not overly important, for example the calculation of the ZPVE or vibrational entropy.
This fitness is in line with the original GFN{\it n}-xTB design purpose\cite{gfnWIREs} and hence seems to be confirmed by our results.
For a visual assessment of \msc, examples of a ``good'', ``average'' and ``mediocre'' matchscore are given in Figure~\ref{fig:spectra_example}, depicted as an overlay between the mass-scaled B3LYP-3c reference and the MACE-OFF23(large)+MACE-$\mu$ computed IR spectra with applied ZPVE scaling factor.
\begin{figure}[ht!]
    \centering
    \includegraphics[width=0.8\linewidth]{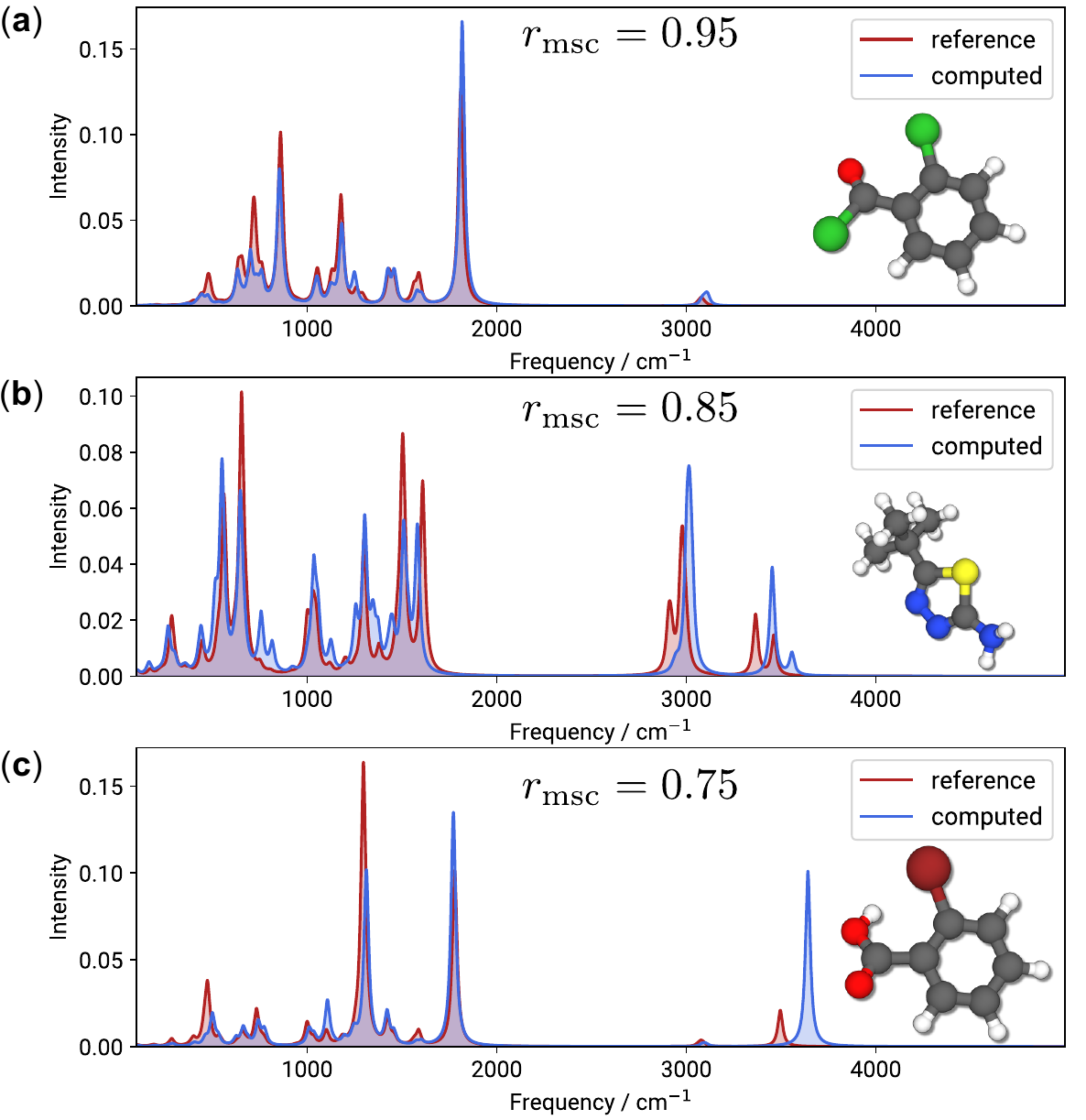}
    \caption{Examples for typical matchscore regimes. \textbf{a}) A ``good'' matchscore case of \msc\,=\,0.95 (NIST Id C609654), \textbf{b}) an ``average'' matchscore case of \msc\,=\,0.85 (NIST Id C39222736), \textbf{c}) a ``mediocre'' matchscore case of \msc\,=\,0.75 (NIST Id C88653). Computed spectra refer to MACE-OFF23(large)+MACE-$\mu$ with frequencies scaled by a factor of 0.9606, reference spectra refer to B3LYP-3c. }
    \label{fig:spectra_example}
\end{figure}
In general, matchscores of 0.9 or higher exhibit a close match between the two compared spectra, with only minor differences in relative intensities. 
Scores between 0.8 and 0.9 likewise denote good correspondence of spectra, although some frequency shifted signals may appear, as seen for modes around 3000 to 3500\,cm$^{-1}$ in Figure.\ref{fig:spectra_example}b.
Spectra with \msc\ below 0.8 commonly disagree with regard to at least one signal, as for the OH vibrational mode at around 3500\,cm$^{-1}$ in Figure~\ref{fig:spectra_example}c, where both the frequency and intensity are overestimated by the low-cost computation.
A mismatch of this type is a common cause of ``mediocre'' or worse matchscore results, despite other regions of the spectrum (e.g. the fingerprint regime up to $\approx$1500\,cm$^{-1}$) exhibiting good agreement. 
Spectra with overall fewer IR signals are more prone to this effect, while signals  in cluttered spectra will often overlap regardless, and hence contribute to \msc.
For benchmark purposes with theory-to-theory comparisons being made, a new metric depending on identification and matching of the actual vibrational mode would clearly be preferable.

\ifx\inclsections\undefined
\else
\section{Conclusion}
\fi
In this work, we tested combinations of low-cost semiempirical, force-field, and MLP methods for calculating gas-phase IR spectra within the doubly-harmonic approximation. 
Our approach involves benchmarking the GFN1-xTB, GFN2-xTB, GFN-FF and MACE-OFF23 models for calculating molecular geometries and associated harmonic frequencies, as well as the GFN1-xTB, GFN2-xTB, CEH, EEQ and MACE-$\mu$ models for molecular dipole moments. 
Our analysis focuses mostly on the IR7193 set, collecting some 7193 organic molecules and their IR spectra.
Linear frequency scaling factors were determined based on the ZPVE fraction relative to a specially (mass-)scaled theoretical reference, B3LYP-3c. 
IR spectra were obtained by combining frequency predictions with IR intensities calculated via numerically derived dipole moment derivatives.

A key observation is that the MACE-OFF23 MLPs are found to be particularly effective for drug-like molecules, providing accurate geometries, frequencies, and IR spectra that outperform the GFN{\it n}-xTB methods. 
Interestingly, MACE-OFF23 appears to inherit characteristics similar to DFT methods, as evidenced by the comparable importance and magnitude of harmonic frequency scaling factors.
This result is contrasts with the evaluated semiempirical methods, which are mostly asystematic with regard to such treatments, and on average are only little affected by a linear scaling of the harmonic frequencies.
In general, although the GFN{\it n}-xTB methods offer decent performance and speed for smaller system sizes and inherently include dipole moments, they often do not match the accuracy of MACE-OFF23. 
This result is clearly seen from the best observed average \msc\ of 0.787 of GFN2-xTB versus a \msc\ of 0.859 for MACE-OFF23(large), both in combination with MACE-$\mu$-based IR intensities.
Both MACE-$\mu$ and GFN{\it n}-xTB exhibit reasonable performance for calculation of the underlying dipole moment derivatives, with the SQM methods having a computational cost advantage for drug-sized molecules.
The use of classical EEQ models for this purpose should be avoided.
While CEH does not reach the accuracy of SQM or MLP dipole predictions, the observed performance is a clear motivator for future research.

In summary, we see an  opportunity for both MLP and SQM-based prediction of IR vibrational spectra, or a combination of techniques.
MACE-OFF23, in particular, provides excellent performance for molecular geometries and frequencies, making it a promising method for calculating thermostatistical contributions for large systems in addition to the spectroscopic applications.
For very large systems, exceeding roughly 700 atoms, we expect MACE-OFF23 to gain a computational cost advantage over the xTB methods due to the MLPs better scaling behaviour with system size. 
Future work should focus on investigating more accurate reference frequencies both by going  beyond the B3LYP-3c level of theory, and extending beyond the harmonic approximation to include vibrational anharmonicity. 
Investigating IR spectra for condensed-phase and large amorphous systems, such as proteins, using MACE-OFF23, is another promising avenue of research.
New MLP developments, especially the inclusion of proper electrostatics,\cite{thomas2024self} will be beneficial for this purpose and could allow IR intensity prediction without the need for separate dipole models.
Additionally, exploring other spectroscopic techniques, such as Raman spectroscopy by examining polarizability instead of dipole moments, will be valuable. 

\clearpage
\subsection*{Glossary}
\begin{table}[ht!]
\scriptsize
\begin{tabular}{l|l}
\hline
\textbf{Abbreviation} & \textbf{Meaning} \\
\hline
BSSE & Basis set superposition error\\
B3LYP-3c &  short notation for B3LYP-D3(BJ)$^\mathrm{ATM}$-gCP/def2-mSVP\\
CEH & Charge extended H\"uckel model\\
CN & Coordination number \\
CREST & Conformer-Rotamer Sampling Tool \\
DFT & Density functional theory \\
DHA & Doubly-harmonic approximation\\
EEQ & Charge equilibrium model\\
FF & Force-field \\
GFN & Geometries, Frequencies, Non-covalent interactions \\
GFN-FF & GFN force-field\\
IR & Infrared (spectroscopy)\\
kQEq & Kernel charge equilibration model\\
MACE & Higher Order Message Passing Atomic Cluster Expansion\\
MACE-OFF23 & MACE organic force-field\\
MACE-$\mu$ & MACE dipole model\\
MAD & Mean absolute deviation \\
MAE & Mean absolute error\\
MD & Mean deviation\\
ML & Machine learning\\
MLP & Machine learning potential \\
PEL & Potential energy landscape (synonym to PES) \\
PES & Potential energy surface (synonym to PEL) \\
RMSD & Root-mean-square-deviation (of atomic positions) \\
RMSE & Root-mean-square-error \\
SPICE & Small-molecule/Protein Interaction Chemical Energies dataset\\
SQM & Semiempirical quantum mechanics \\
xTB & eXtended Tight-Binding \\
ZPVE & Zero-point vibrational energy \\
\hline
\end{tabular}
\label{table:glossary}
\end{table}

\subsection*{Acknowledgement}
This research was supported by the Machine Learning for Pharmaceutical Discovery and Synthesis consortium.
P.P. gratefully acknowledges support by the Alexander von Humboldt Foundation for a Feodor Lynen Research Fellowship and by the DAAD for a postdoctoral research fellowship.
V.K. acknowledges support from the startup fund provided by the Department of Physics and Astronomy, University College London and is grateful for computational resources from UCL and the Swiss National Supercomputing Centre under project s1281.
Y.P. acknowledges the support of Cambridge Trust for awarding a Cambridge international scholarship. 
The authors thank Marcel M\"uller, Stefan Grimme, and William\,H.\,Green for discussions and support.



\bibliography{lit}

\end{document}


\maketitle
\newpage

\section{Calculation of spectra comparison scores}

In computed spectra the frequencies and intensities 
of the vibrational modes are available as isolated signals ('\textit{stick spectrum}') and for detailed comparisons have to be expanded into the same spectral domain as, e.g. experimental data.
This is achieved by employing a Lorentzian line shape function for each mode 
\begin{align}
\phi_p(\nu) = I_p \left( 1 + \frac{\nu_p - \nu }{0.5w}  \right)^{-1} \,\, ,
\end{align}
where $\nu_p$ is the position (calculated frequency) of the mode $p$, $I_p$ is its intensity and $w$ is the full width at half maximum (FWHM).
Typical values employed for the FWHM in range from 20 to 40\,cm$^{-1}$, whereas the average line width in experimental spectra was determined as 24\,cm$^{-1}$.\cite{chu1999} 
The simulated spectrum is then simply given by the sum of all the Lorentzian functions over all $N_p$ modes
\begin{align}
\Phi_ \mathrm{norm}(\nu) = I_\mathrm{norm}\sum_p^{N_p} \phi_p(\nu) \,\, ,
\label{qs:sumphi}
\end{align}
with the normalization constant $I_\mathrm{norm}$.
Such spectra are normalized by $\sqrt{\int\Phi_\mathrm{norm}\mathrm{d}\nu}\overset{!}{=}1$ and in this form are directly
(point wise) comparable to experimental and other computed data. 
Normalizations to single signals, e.g., the largest peak of the spectrum, should be avoided because relative intensities and frequencies are strongly dependent on the theoretical level and hence it cannot always be ensured that the same peak is selected.
Finally, from the two normalized spectra that shall be compared, two $k$-dimensional vectors ($u$ and $v$) are constructed where $u_i$ is the normalized intensity of the $i$-th point in the spectrum and $\mathrm{d}u$ is a predefined distance resolution between the points $u_i$ and $u_{i+1}$. 
In summary, the FWHM, $k$ and $\mathrm{d}u$ are input parameters affecting the comparison.
For consistency, all comparisons employed the same defaults with $\mathrm{d}u$ set to 1.0\,cm$^{-1}$), and $k$ chosen accordingly in the regime between 100 and 4500 cm$^{-1}$.
For the FWHM a value of 30\,cm$^{-1}$ is selected, which is physically realistic but still provides some leeway for frequency deviations between the spectra.

Four different spectral similarity measures have been employed in accordance with previous studies.\cite{baumann1997,song2017,petrich2011,zapata2018,pracht2020_1} 
The first is a simple match score (\msc),
\begin{align}
\text{\msc} = \frac{\left( \sum^k_{i}u_{i}v_{i} \right)^2}{\left(\sum^k_{i} u_{i}^2\right) \left(\sum^k_{i}v_{i}^2 \right)} \,\, ,
\label{eq:MSC}
\end{align}
where $u$ and $v$ are the $k$-dimensional vectors obtained for the two compared spectra.
The \msc\  corresponds to a Cauchy-Schwarz inequality in $\mathbb{R}^k$ dimensional Euclidian space which essentially is
a simplified overlap. \msc\  values range from $0 \leq \text{\msc} \leq 1$, where unity denotes a perfect match.
If not stated otherwise, the \msc\ is our default unit of measurement for spectral comparison.

The second measure used is the Euclidean norm (\euc),
\begin{align}
\text{\euc} = \left( 1.0 + \frac{\sum_i^k(u_i- v_i)^2}{\sum_i^k(v_i)^2} \right)^{-1} \,\, .
\label{eq:EUC}
\end{align}

The third measure is the Pearson correlation coefficient (\pcc), which is similar to the \msc,\begin{align}
\text{\pcc} = \frac{\sum_{i}^k(u_{i}-\bar{u})(v_{i} -\bar{v})}{\sqrt{\sum_{i}^k(u_{i}-\bar{u})^2}\sqrt{\sum_{i}^k(v_{i}-\bar{v})^2}} \,\, ,
\label{eq:PCC}
\end{align}
with the mean values $\bar{u}$ and $\bar{v}$ for $u$ and $v$. Both the \msc\  and \pcc\  are linear correlation measures that are based on the Cauchy-Schwarz inequality.
And finally, the fourth measure is the Spearman rank correlation coefficient (\scc)
\begin{align}
\text{\scc} = 1.0 - \frac{6\sum_i^k(rg(u_i)- rg(v_i))^2}{k\left(k^2 - 1 \right)} \,\, ,
\label{eq:SCC}
\end{align}
where $rg(u_i)$ and $rg(v_i)$ are the respective ranks of $u_i$ and $v_i$.

In the literature, other measures have been proposed and adapted for various purposes,\cite{penchev1996} however, in the context of spectra comparison, Henchel et al.\cite{henschel2020} suggested that the different similarity scores have different advantages and disadvantages and can be used complementary, which is in line with our own previous work.\cite{pracht2020_1}

\section{Dataset composition}
\begin{figure}[ht!]
    \centering
    \includegraphics[width=1\linewidth]{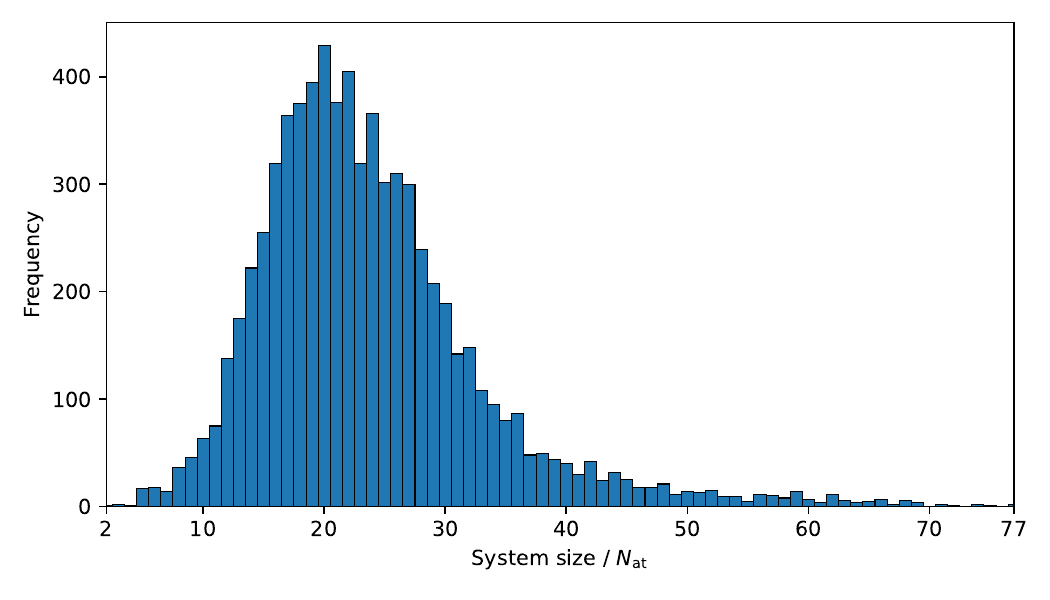}
    \caption{System sizes in IR7193.}
    \label{fig:syssize}
\end{figure}

\begin{figure}[ht!]
    \centering
    \includegraphics[width=1\linewidth]{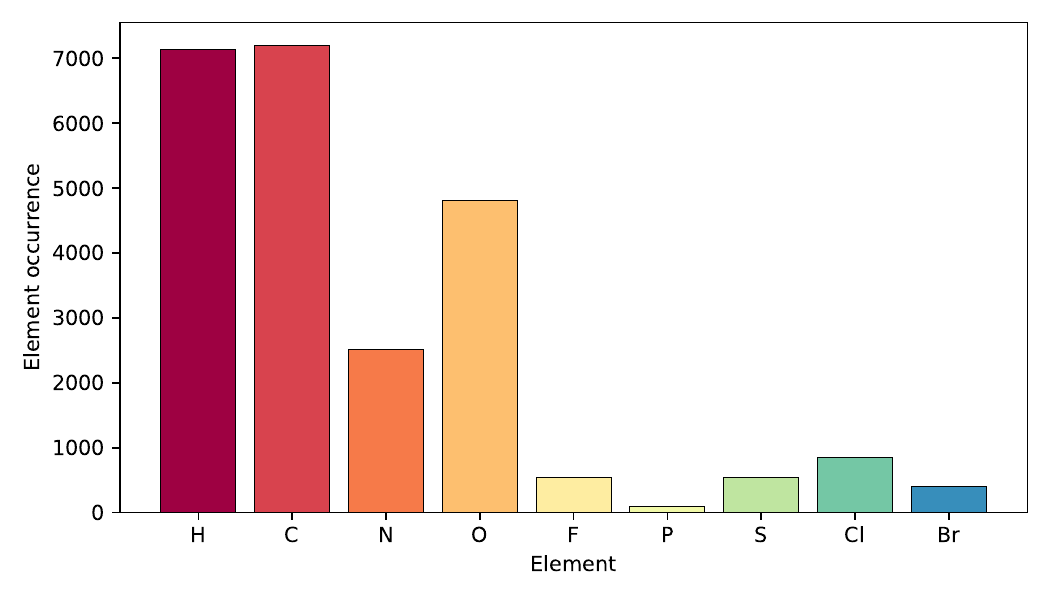}
    \caption{Element occurrence histogram for the IR7193 database.}
    \label{fig:elements}
\end{figure}

The IR7193 dataset was originally\cite{pracht2020_1} proposed by obtaining all experimentally available gas-phase IR spectra of the NIST database.\cite{nist}
The distribution of system sizes in IR7193 is shown in Figure~\ref{fig:syssize}.
The occurrence of different elements is visualized in Figure~\ref{fig:elements}. Systems with the elements HCNO are present in abundance. The smallest subset are phosphorous containing systems with exactly 100 molecules.

\section{Molecular geometries at $\omega$B97M-D3(BJ) level}

As an additional perspective on the molecular geometry quality, all structures of IR7913 were reoptimized at the $\omega$B97M-D3(BJ)/def2-TZVPP level of theory.
This reference level only differs from the SPICE set\cite{Eastman2023} reference by the additional diffuse functions in the basis set, which were omitted here for sake of computational performance.
However, according to best DFT practices,\cite{dft_best_practice} it is reasonable to assume that differences between the two basis sets are insignificant for optimized geometries.
Possibly much greater differences can be expected between B3LYP-3c and $\omega$B97M-D3(BJ)/def2-TZVPP.
Despite employing some hybrid-level DFT in both cases, the basis sets differ in cardinal number, which is a major factor for both computational cost and performance. 
The range-separation of exchange in $\omega$B97M-D3 further represents a key difference which influences the molecular geometries.

\begin{figure}[ht!]
    \centering
    \includegraphics[width=0.99\linewidth]{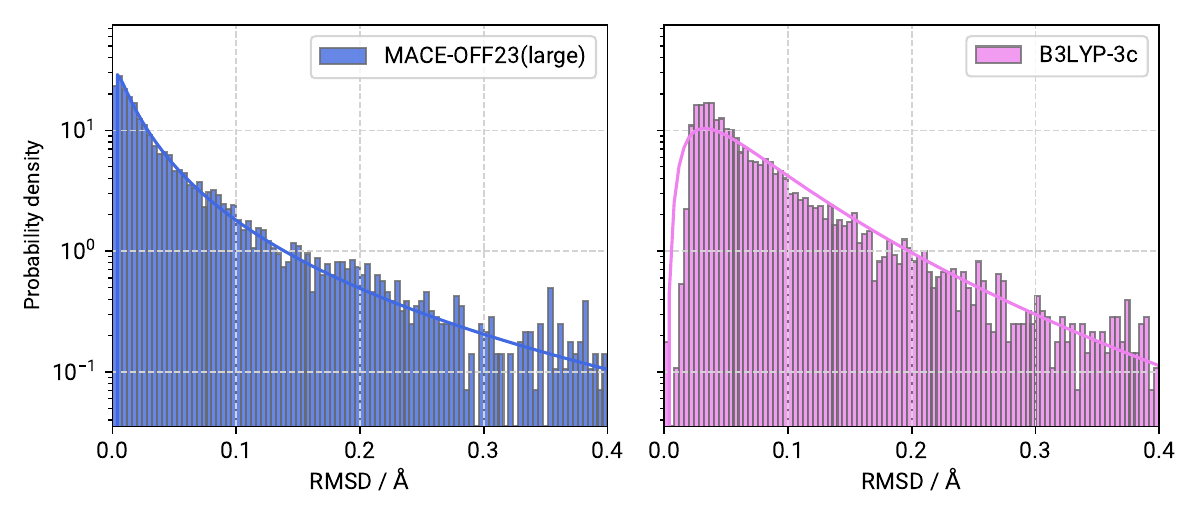}
    \caption{Histograms and fitted log-normal distributions for Cartesian RMSDs calculated between the $\omega$B97M-D3(BJ)/def2-TZVPP reference and the MACE-OFF23(large) and B3LYP-3c minima for the IR7193 set. All plots use a logarithmic scale to emphasize the distribution tails.}
    \label{fig:si-rmsds1}
\end{figure}

\begin{figure}[ht!]
    \centering
    \includegraphics[width=0.99\linewidth]{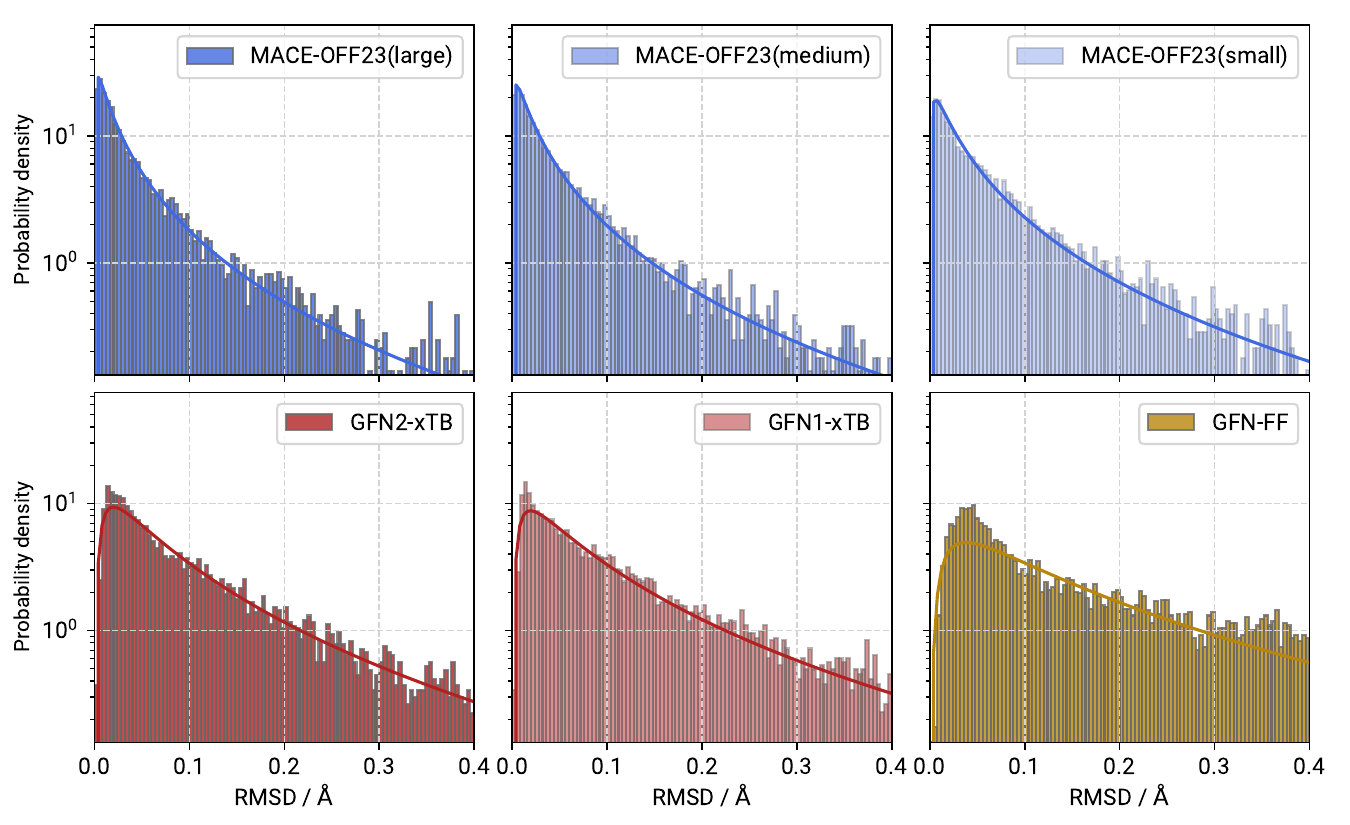}
    \caption{Histograms and fitted log-normal distributions for Cartesian RMSDs calculated between the $\omega$B97M-D3(BJ)/def2-TZVPP minima and corresponding optimized molecular structures at B3LYP-3c, MACE-OFF23(small/large) and GFN{\it n}-xTB/FF levels of theory for the IR7193 set. All plots use a logarithmic scale to emphasize the distribution tails.}
    \label{fig:si-rmsds2}
\end{figure}

A comparison of the performance of B3LYP-3c and MACE-OFF23(large) is shown in Figure~\ref{fig:si-rmsds1}. The full evaluation of the MACE and xTB low-cost potentials, analogous to the main article, can be found in Tab.~\ref{tab:si-rmsds}, Tab.~\ref{tab:si-rmsdspercent} and Figure~\ref{fig:si-rmsds2}.
\begin{table}
    \centering
    \begin{adjustbox}{center}
        \begin{tabular}{l|c|ccc|ccc}
          & \textbf{B3LYP-3c} & \multicolumn{3}{c|}{\textbf{MACE-OFF23 model}} & \textbf{GFN2-xTB} & \textbf{GFN1-xTB} & \textbf{GFN-FF}\\
           & & \textbf{small} & \textbf{medium} & \textbf{large} &&& \\ \hline
    \textbf{Mean}	& 0.0951 & 0.0878 & 0.0704 & 0.0642 & 0.1351 & 0.1504 & 0.2419\\
    \textbf{Median}	& 0.0571 & 0.0419 & 0.0308 & 0.0262 & 0.0686 & 0.0800 & 0.1369\\
    \textbf{SD}	    & 0.1177 & 0.1379 & 0.1182 & 0.1141 & 0.1864 & 0.2060 & 0.2668\\
    \end{tabular}
    \end{adjustbox}
    \caption{Mean, median, and standard deviation (SD) for Cartesian RMSDs calculated between the $\omega$B97M-D3(BJ)/def2-TZVPP reference and B3LYP-3c, MACE-OFF23(small/medium/large) and GFN{\it n}-xTB/FF optimized structures of IR7193. All values are in {\AA}ngstr\"om. Narrow distributions indicate better performance.}
    \label{tab:si-rmsds}
\end{table}
\begin{table}[ht!]
    \centering
    \begin{adjustbox}{center}
\begin{tabular}{c|c|ccc|ccc}
     \textbf{RMSD} & \textbf{B3LYP-3c} & \multicolumn{3}{c|}{\textbf{MACE-OFF23 model}} & \textbf{GFN2-xTB} & \textbf{GFN1-xTB} & \textbf{GFN-FF}\\
    \textbf{[{\AA}]} && \textbf{small} & \textbf{medium} & \textbf{large} &&& \\ \hline
$\leq$\,0.2	& 90.31\% & 88.84\% & 91.77\% & 92.88\% & 80.88\% & 77.78\% & 60.02\% \\
$\geq$\,0.5	& 1.46\% & 2.03\% & 1.33\% & 1.24\% & 5.24\% & 6.02\% & 14.90\% \\
$\geq$\,1.0	& 0.26\% & 0.35\% & 0.22\% & 0.24\% & 0.71\% & 1.13\% & 2.14\% \\
\end{tabular}
    \end{adjustbox}
    \caption{Percentage of structures for IR7193 falling within the specified Cartesian RMSD threshold at a given level of theory.}
    \label{tab:si-rmsdspercent}
\end{table}

Two main effects are observed when referring to $\omega$B97M-D3(BJ)/def2-TZVPP minima instead of B3LYP-3c in Figure~\ref{fig:si-rmsds2}: 
First, the log-normal distribution peak is notably left-shifted for all low-cost potentials, and secondly, the distribution tails are longer in each case.
This is reflected by both table \ref{tab:si-rmsds} and \ref{tab:si-rmsdspercent}, in particular for the MACE-OFF23 models which show a similar a mean, lower median and higher standard deviation compared to the B3LYP-3c-based results from the main article.
Here, while cases with RMSDs below 0.2\,\AA{} are fewer compared to the previous comparison, the percentage of even lower RMSD pairs \textit{within} the 0.2\,\AA{} window increased leading to the left-shifted log-normal distribution maxima.
Since MACE-OFF23 is trained to $\omega$B97M-D3(BJ) data via the SPICE set, this performance is not unexpected although the overall errors with regard to either DFT reference seems small.
GFN{\it n}-xTB/FF, on the other hand, show slightly larger disagreements with $\omega$B97M-D3(BJ)/def2-TZVPP compared to the previously evaluated B3LYP-3c. 
A possible explanation, in particular for the xTB methods, is that these share some similarities with the "3c" method family and have been parametrized to the PBEh-3c\cite{pbeh3c} and B97-3c\cite{b973c} data.\cite{gfnWIREs}
Finally, comparing B3LYP-3c and $\omega$B97M-D3(BJ)/def2-TZVPP, one can observe some performance differences between the two DFT methods. 
Overall, B3LYP-3c performs \textit{worse} than MACE-OFF23(medium/large) which becomes apparent from Figure~\ref{fig:si-rmsds1}, but better than GFN{\it n}-xTB/FF and MACE-OFF23(small). 
As noted above, we expect these differences to be a result of the much larger basis set and the range-separated exchange present in $\omega$B97M-D3(BJ)/def2-TZVPP, making the latter a generally better choice for a reference level of theory.
Unfortunately, the large difference in computational cost between the two DFT methods currently prevents us from calculating frequencies at the $\omega$B97M-D3(BJ) level at present time, as this goes much beyond the task of obtaining molecular geometries. We are planning a revision for future work.

\clearpage
\section{Dipole moment benchmarking}
\label{sec:si-dipole}

Our reference method for dipole moments in the main manuscript is $\omega$B97M-D3(BJ)/def2-TZVPPD, since MACE is trained to this level of theory via the SPICE dataset.\cite{Eastman2023}
For a very few number of cases the smaller basis set def2-TZVPP had to be employed to circumvent SCF convergence issues. 
In total, this affected 42 out of 7193 molecules (0.58\,\%).

\begin{figure}[ht!]
    \centering
    \includegraphics[width=1\linewidth]{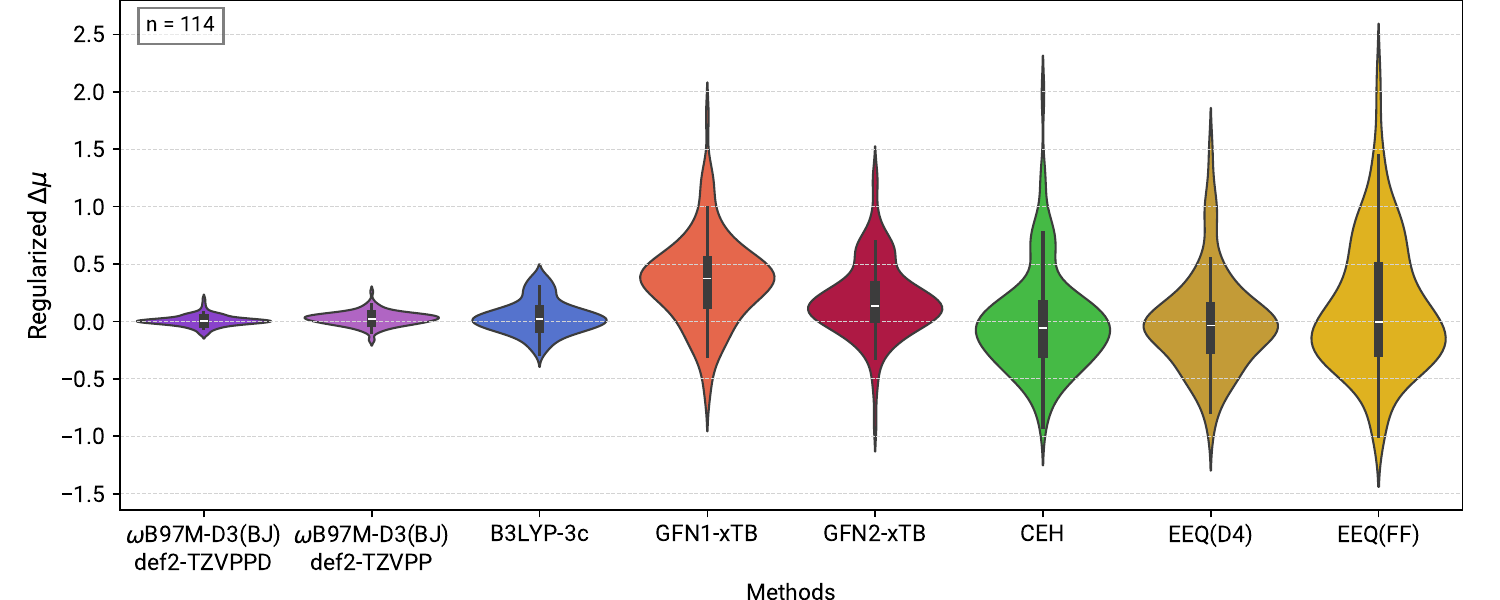}
    \caption{Violin plots for regularized dipole differences $\Delta \mu$ obtained from deviations between reference CCSD(T)/CBS dipoles of the MHG114 set\cite{mhg2018,Zapata2020} and dipole moments calculated at the shown levels of theory.}
    \label{fig:mhg114}
\end{figure}

To estimate the performance of these reference methods to even high-level reference data, a benchmark set of CCSD(T)/CBS extrapolated dipole moments of small molecules was used.
The original set was published by Head-Gordon and coworkers,\cite{mhg2018} however, we refer to a slightly smaller subset of 114 molecules which excludes transition metals, as used by Zapata et al.\cite{Zapata2020}
This benchmark set is abbreviated as MHG114 in the following.
The respective data is shown in Figure~\ref{fig:mhg114} and Tables~\ref{tab:mhg114_1} and \ref{tab:mhg114_2}.

\begin{table}[ht!]
    \centering
\begin{tabularx}{\textwidth}{XXXXXX}\hline
 & $\omega$B97M-D3(BJ)/def2-TZVPPD & $\omega$B97M-D3(BJ)/def2-TZVPP & B3LYP-3c & GFN1-xTB & GFN2-xTB \\ \hline
MD & 0.0051 & 0.0238 & 0.0383 & 0.3553 & 0.1881  \\
MAD & 0.0327 & 0.0449 & 0.1087 & 0.2752 & 0.2338  \\
RMSE & 0.0475 & 0.0609 & 0.1429 & 0.3789 & 0.3209  \\
\hline
\end{tabularx}
    \caption{Errors in units of regularized dipole moment deviations for DFT and SQM methods tested on the MHG114 set of CCSD(T)/CBS dipole moments.}
    \label{tab:mhg114_1}
\end{table}

\begin{table}
    \centering
\begin{tabularx}{\textwidth}{XXXXXX}\hline
 & CEH & EEQ(D4) & EEQ(GFN-FF) & MACE-$\mu$ medium & MACE-$\mu$ small\\ \hline
MD &  -0.0127 & -0.0157 & 0.1266 & -- & --\\
MAD &  0.3033 & 0.2932 & 0.4470 & -- & --\\
RMSE &  0.4294 & 0.4178 & 0.5676 & -- & --\\
\hline
\end{tabularx}
    \caption{Errors in units of regularized dipole moment deviations for (semi-)classical and ML methods tested on the MHG114 set of CCSD(T)/CBS dipole moments.}
    \label{tab:mhg114_2}
\end{table}

Overall, a clear correlation with the level of theory can be observed for the tested methods can be observed.
With an $\Delta \mu$ MAD of only 3.27\%, $\omega$B97M-D3(BJ)/def2-TZVPPD shows excellent performance on the MHG114 set and surely is sufficient to serve as our reference for further comparisons.
The basis set size slightly affects (worsens) this performance, shown through $\omega$B97M-D3(BJ)/def2-TZVPP with an MAD of 4.49\%.
B3LYP-3c, with an MAD of 10.87\%, provides a balanced performance that is in line with previous observations for smaller basis sets.\cite{Zapata2020}
The semiempirical GFN1-xTB and GFN2-xTB methods, as well as the semi-classical models (Table~\ref{tab:mhg114_2}) show much larger deficiencies.
Partly, this can be attributed to a number of open-shell systems in the benchmark set, which are either insufficiently or entirely incorrectly described at these levels of theory. 
This observation was already made during the GFN2-xTB method development.\cite{gfn2xtb}
The MACE-$\mu$ models were not evaluated on the MHG114 set since several open-shell molecules are contained within and there is no option for treating such systems with the MLP.

\begin{figure}[ht!]
    \centering
    \rotatebox{90}{\includegraphics[width=1.3\linewidth]{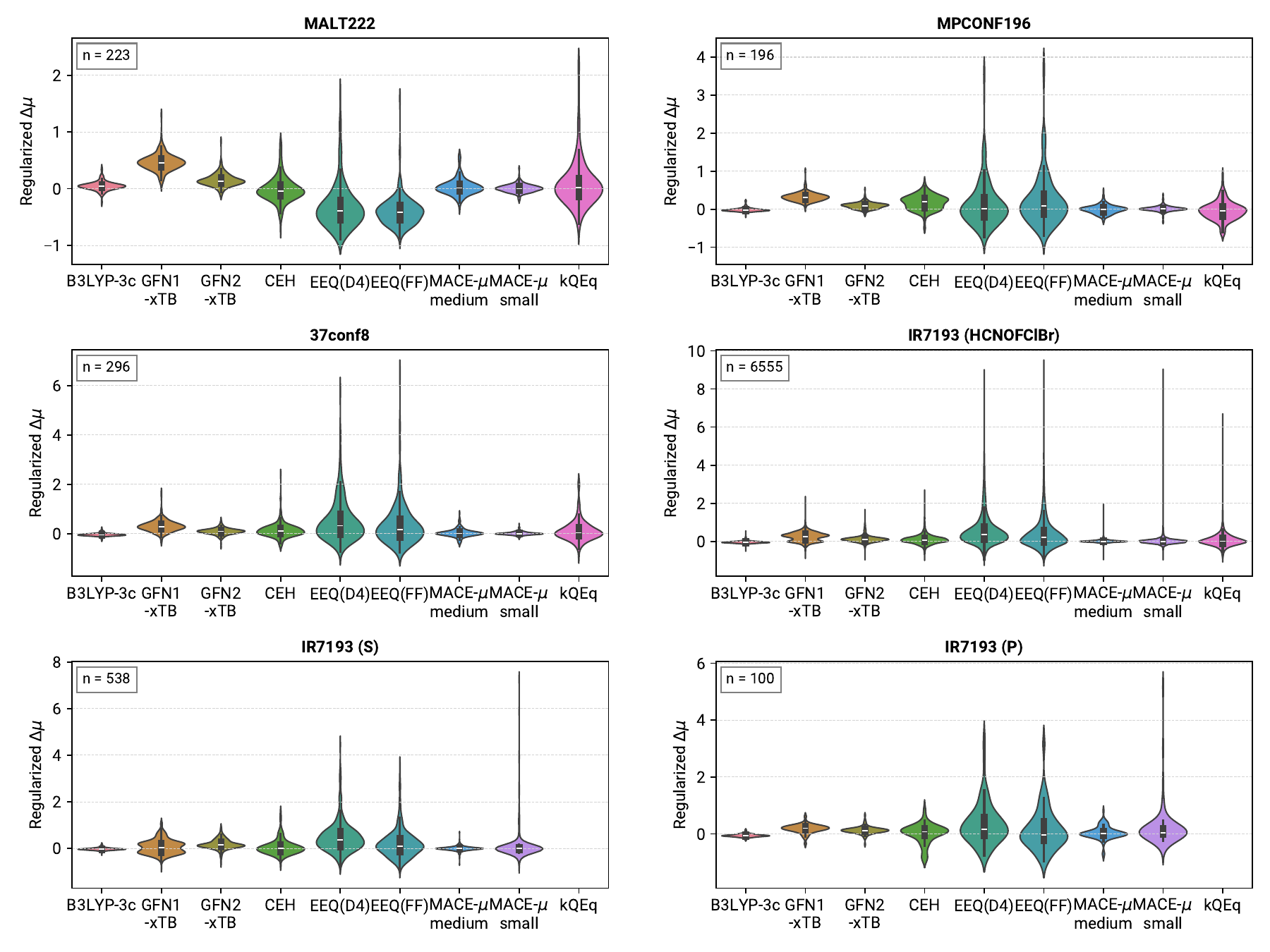}}
    \caption{Violin plots for regularized dipole moment differences $\Delta \mu$, shown for a variety of benchmark subsets and methods. Reference dipole moments were obtained at the $\omega$B97M-D3(BJ)/def2-TZVPPD level of theory. Figure rotated by 90$^\circ$ to fit landscape page view.}
    \label{fig:dipolebenchmarks}
\end{figure}

Since molecules in the MHG114 are small and far from typical ``use-case'' molecules (e.g. drugs), we benchmarked low-cost methods on a far larger set of structures, taking $\omega$B97M-D3(BJ) dipole moments as a reference.
The benchmark set was composed out of the IR7193 set of molecules, as well as three smaller benchmark sets focusing on different conformations of molecules: MALT222,\cite{glucose-maltose} MPCONF196\cite{rezac2018}, and 37conf8.\cite{shapara2019}
The summary of the evaluations are given in the main manuscript. 
However, violin plots for regularize dipole moment differences of each individual set can be found in Figure~\ref{fig:dipolebenchmarks}.

\clearpage
\subsection{Charge distribution of dodecanoic acid}
\begin{figure}
    \centering
    \includegraphics[width=0.85\linewidth]{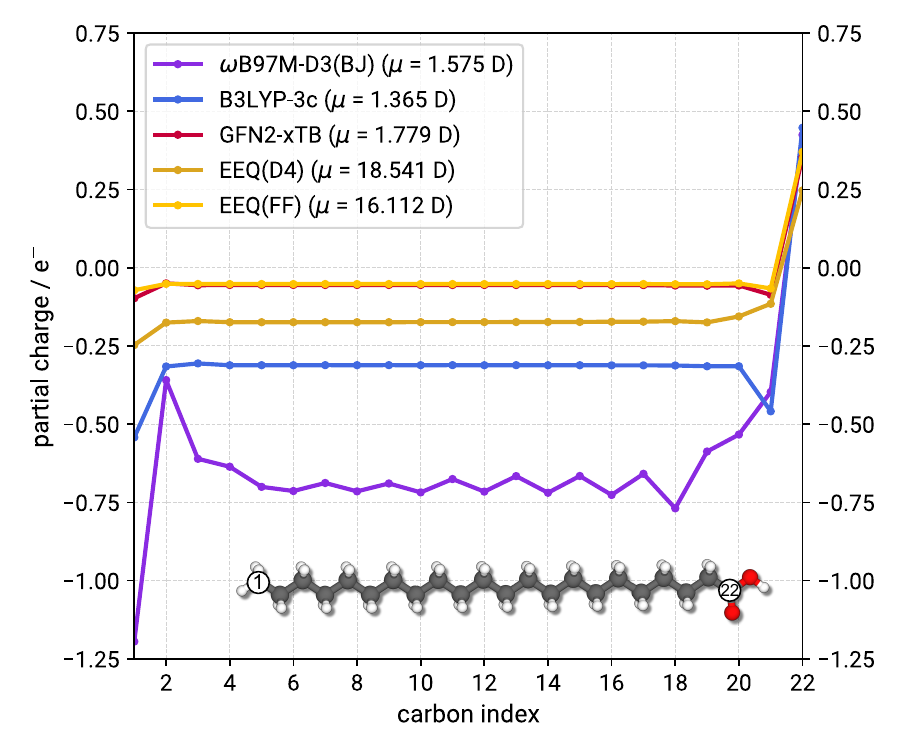}
    \caption{Distribution of partial charges on the carbon atoms of dodecaonic acid, shown for \refwBdef, B3LYP-3c, GFN2-xTB, EEQ(D4) and EEQ(GFN-FF). Carbon atoms are numbered sequentially 1 to 22, as shown.}
    \label{fig:partial_e}
\end{figure}

\subsection{MACE-$\mu$ model training}
Fitting dipoles within the MACE architecture only requires implementing the readout with L=1 equivariance. We use 128 equivariant channels with a maximum equivariant order of L=1 and a cutoff of 6\,\AA{} with two interaction layers, resulting in a receptive field of 12\,\AA{}. The readout multi layer perceptron uses 16 equivariant channels. We train directly on the total dipole moments, without using any point charge baseline, on the same dataset\cite{Eastman2023} as the MACE-OFF series of models.

\clearpage
\section{Timings}
\subsection{Timings for potential evaluations}

\begin{figure}[ht!]
    \centering
    \includegraphics[width=1.0\linewidth]{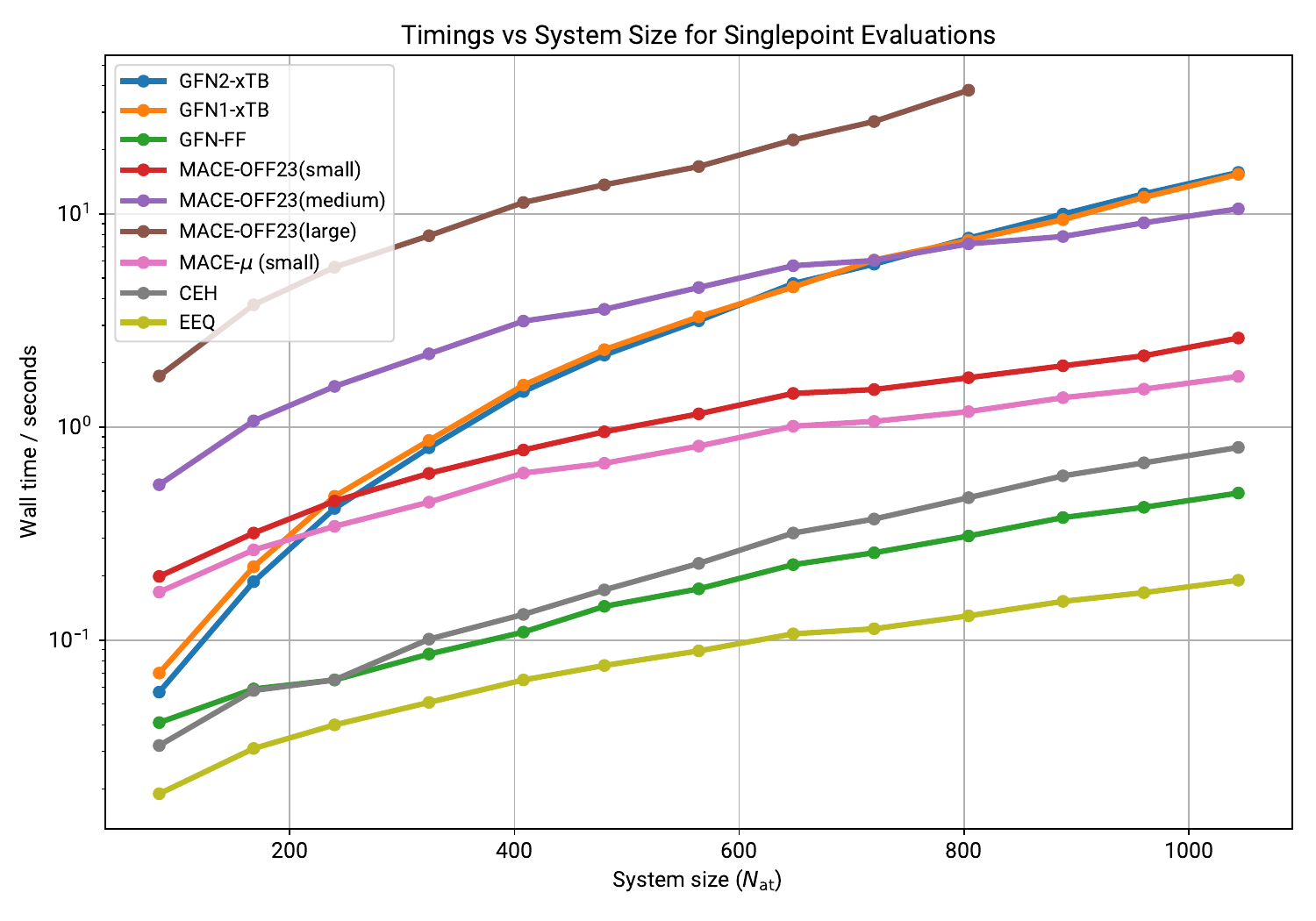}
    \caption{Singlepoint evaluation wall-times vs. system size for methods employed in this study. Note a logarithmic scale is used. The smallest system has 84 atoms, the largest system has 1044 atoms. All calculations refer to CPU timings and use identical hardware.}
    \label{fig:scalingsp}
\end{figure}

Computational wall-times for simple potential evaluation were compared for thirteen model systems ranging in size from 84 to 1044 atoms.
The model systems correspond to non-covalent clusters modelling the Miller-Urey experiment, all with a 1:1:1 ratio of H$_\mathrm{2}$O, NH$_\mathrm{3}$ and CH$_\mathrm{4}$ molecules.
All calculations were performed using a 11th Gen Intel Core i7-11800H (2.30GHz) processor and 4 shared memory threads.
The documented timings are shown in Figure~\ref{fig:scalingsp}.
Calculations for the MACE-OFF23(large) for systems with 888, 960, and 1044 atoms terminated early because of extensive memory requirements, exceeding the available 16\,GB RAM plus 7\,GB swapping partition.
MACE-OFF23 timings do \textit{not} include model initialization time. Note, even further speed-up for the MACE-OFF23 models is possible by utilizing the GPU implementation. The CPU implementation was used here to ensure direct comparability to the xTB methods.
All calculations were repeated 3 times to obtain an average wall-time for each corresponding system size.

\clearpage
\subsection{Timings for Hessian evaluations}

Computational wall-times were investigated for a subset of IR7193, taking one molecule for each available size (2 to 77 atoms).
To highlight the performance for MACE-OF23, wall-times were documented for these molecules using both the numerical Hessian implementation, and the ``autograd'' implementation.
All calculations were performed using a 11th Gen Intel Core i7-11800H (2.30GHz) processor and a variable number of shared memory threads.

\begin{figure}[ht!]
    \centering
    \includegraphics[width=0.9\linewidth]{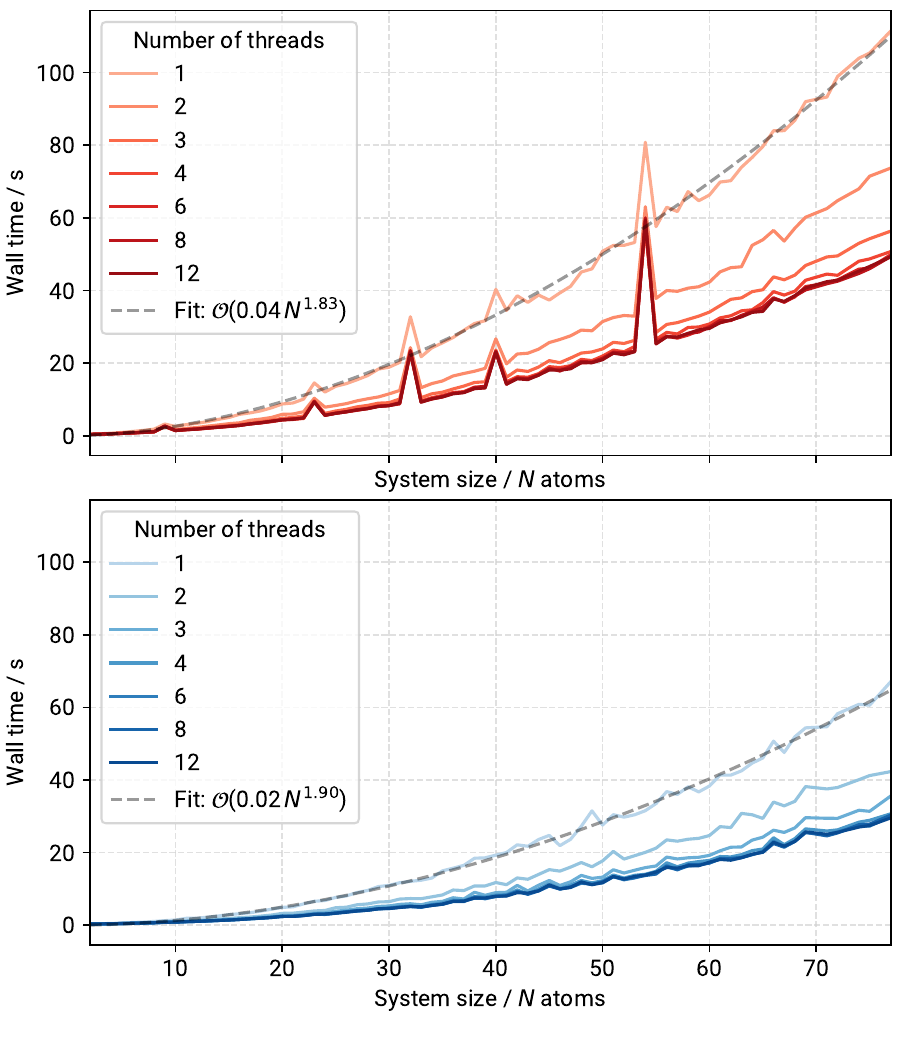}
    \caption{Shared memory parallelization (OpenMP) acceleration for numerical (top, red) and autograd (bottom, blue) Hessian calculations with the MACE-OFF23(small) model. Obtained on a 11th Gen Intel Core i7-11800H (2.30GHz) processor.}
    \label{fig:scalingomp}
\end{figure}

Expectantly, Hessian calculation wall-times should scale quadratically with the system size $N$, the molecule's number of atoms.
Generalizing this relation for comparison, a simple Power-law scaling function $\mathcal{O}(aN^b)$ was fitted to the documented wall-times.
The determined Power-law factors $a$ and $b$ for the small, medium, and large MACE models are shown in Table~\ref{tab:scaling}, parallelization was tested and is shown only for the MACE-OFF23(small) model in Figure~\ref{fig:scalingomp}.

\begin{table}
    \centering
    \begin{tabularx}{0.9\textwidth}{lXZZ}
        \textbf{MACE-OFF23 model} & \textbf{Hessian type} & $a$ & $b$ \\ \hline
        small 	& numerical 	& 0.0216 	& 1.8108 \\
        medium 	& numerical 	& 0.0394 	& 2.0962 \\
        large 	& numerical 	& 0.1614 	& 2.0161 \\
        small 	& autograd 	& 0.0047 	& 2.0469 \\
        medium 	& autograd 	& 0.0224 	& 2.1469 \\
        large 	& autograd 	& 0.1441 	& 1.9613 \\\hline
    \end{tabularx}
    \caption{Fitted factors $a$ and $b$ of the Power-law scaling function $\mathcal{O}(aN^b)$ for the specified MACE-OFF23 and Hessian calculation type combinations. The data corresponds to timings given in Figure~4 in the main article. $N$ is the system size (number of atoms), all calculations used 4 threads for shared memory (OpenMP) parallelization. Obtained on a 11th Gen Intel Core i7-11800H (2.30GHz) processor.}
    \label{tab:scaling}
\end{table}

\clearpage
\bibliography{lit}